\documentstyle[12pt]{article}
\begin{document}
\makeatletter
\def\siml{\mathrel{\mathpalette\gl@align<}}
\def\simg{\mathrel{\mathpalette\gl@align>}}
\def\gl@align#1#2{\lower.6ex\vbox{\baselineskip\z@skip\lineskip\z@
 \ialign{$\m@th#1\hfill##\hfil$\crcr#2\crcr{\sim}\crcr}}}
\makeatother
\hbadness=10000
\hbadness=10000
\begin{titlepage}
\nopagebreak
\def\thefootnote{\fnsymbol{footnote}}
\begin{flushright}
{\normalsize
 DPSU-96-11\\
INS-Rep-1153\\
August, 1996   }\\
\end{flushright}
\vspace{1cm}
\begin{center}
\renewcommand{\thefootnote}{\fnsymbol{footnote}}
{\large \bf  Generic Formula of Soft Scalar Masses\\
 in String Models}

\vspace{1cm}

{\bf Yoshiharu Kawamura $^a$ 
\footnote[1]{e-mail: ykawamu@gipac.shinshu-u.ac.jp}}
and  \ {\bf Tatsuo Kobayashi $^b$
\footnote[2]{e-mail: kobayast@ins.u-tokyo.ac.jp}}

\vspace{1cm}
$^a$ Department of Physics, Shinshu University \\

   Matsumoto, 390 Japan \\
and\\
$^b$ Institute for Nuclear Study, University of Tokyo \\
   Midori-cho, Tanashi, Tokyo, 188 Japan \\

\end{center}
\vspace{1cm}

\nopagebreak

\begin{abstract}
We derive formula of soft supersymmetry breaking scalar masses
from 4-dimensional string models within a more generic framework.
We consider effects of extra gauge symmetry breakings
including an anomalous $U(1)$ breaking through flat directions,
that is, $D$-term and $F$-term contributions, particle mixing
effects and heavy-light mass mixing effects.
Some phenomenological implications are discussed based on our 
mass formula.
\end{abstract}
\vfill
\end{titlepage}
\pagestyle{plain}
\newpage
\section{Introduction}
\renewcommand{\thefootnote}{\fnsymbol{footnote}}

Superstring theories (SSTs) are powerful candidates for the unification
theory of all forces including gravity.
There are various approaches to explore 4-dimensional (4-D) string models, 
for example, the compactification on Calabi-Yau manifolds \cite{CY}, 
the construction of orbifold models \cite{Orb,4DST} and so on.
The effective supergravity theories (SUGRAs) have been derived
based on the above approaches \cite{ST-SG,OrbSG,OrbSG2}.
The structure of SUGRA \cite{SUGRA} is constrained
by considering field theoretical non-perturbative effects 
such as a gaugino condensation \cite{gaugino} 
and stringy symmetries such as duality \cite{duality}
besides of perturbative results.

Effective theories, however, have several problems.
First, there are thousands of effective theories corresponding to
4-D string models.
They have, in general, large gauge groups and many matter multiplets
compared with those of the minimal supersymmetric standard 
model (MSSM).
We have not known how to select a realistic model among them
from stringy theoretical point of view yet.
Another serious problem is that the mechanism of supersymmetry 
(SUSY) breaking is unknown.
To solve these problems, non-perturbative effects
in SSTs and SUSY field theories should be fully understood.\footnote{
Recently there have been various remarkable developments 
in study on non-perturbative aspects of SSTs and SUSY models
\cite{STnp,SUSYdual}.}

At the present circumstance, the following approaches and/or standpoints
have been taken.
For the first problem, study on flat directions 
is important \cite{flat1}.
Because effective theories have, in general, flat directions 
in the SUSY limit.
Large gauge symmetries can break into
smaller ones and extra matter fields can get massive
through flat directions.
Further flat directions could relate different models in string vacua.
Actually some models with realistic gauge groups and matter contents 
have been constructed based on $Z_3$ orbifold models \cite{flat2}.
Recently generic features of flat directions in $Z_{2n}$ orbifold models 
have been also investigated \cite{flat3}.

The flat directions based on $Z_3$ orbifold models have been analyzed
considering the existence of anomalous $U(1)$ symmetry $(U(1)_A)$
because 4-D string models, in general, have the $U(1)_A$ symmetry.
Some interesting features are pointed out in those models.
For example, Fayet-Iliopoulos $D$-term \cite{FI} is induced 
at one-loop level
for $U(1)_A$ \cite{ST-FI}.\footnote{
Some conditions for absence of anomalous $U(1)$ are 
discussed in ref.\cite{anom}.}
As a result, some scalar fields necessarily develop vacuum expectation 
values(VEVs) and some gauge symmetries can 
break down \cite{flat1,flat2}.

For the second problem,
some researches have been done from the standpoint that the origin of 
SUSY breaking is unspecified.
That is, soft SUSY breaking terms have been 
derived under the assumption that
SUSY is broken by $F$-term condensations of the dilaton field $S$
and/or moduli fields $T$ \cite{IL,ST-soft,ST-soft2}.
Some phenomenologically interesting features are predicted from 
the structure of soft SUSY breaking terms which
are parameterized by a few number of parameters, for example,
only two parameters such as a goldstino angle $\theta$ and the gravitino 
mass $m_{3/2}$ in the case with the overall moduli and the vanishing 
vacuum energy \cite{BIM}.
The cases with multimoduli fields are also discussed 
in Refs.\cite{multiT}.
Recently study on soft scalar masses has been extended
in the presence of an anomalous $U(1)$ symmetry \cite{N,U1A,U1X}.

This strategy for string phenomenology is quite interesting
since the soft SUSY breaking parameters can be
powerful probes for physics beyond the MSSM such as SUSY-grand unified
theories (SUSY-GUTs), SUGRAs and SSTs.
We give two examples.
The pattern of gauge symmetry breakdown can be specified
by checking certain sum rules among scalar masses.
The specific mass relations are derived 
for $SO(10)$ breakings \cite{KMY1,KMY2} and 
for $E_6$ breakings \cite{KT}.
String models with the SUSY breaking due to the dilaton
$F$-term lead to the highly restricted pattern 
such as \cite{ST-soft,BIM}
\begin{eqnarray}
|A| = |M_{1/2}| = \sqrt{3} |m_{3/2}| 
\label{dilaton}
\end{eqnarray}
where $A$ is a universal $A$-parameter, and gauginos and scalars get masses 
with common values $M_{1/2}$ and $m_{3/2}$, respectively.
In this way, soft SUSY breaking parameters can play important 
roles to probe a new physics.

The above two approaches are attractive to explore particle phenomenology
beyond the standard model based on SST.
Hence it is important to examine what features soft 
SUSY breaking terms can show at low energy when 
we construct a realistic model through flat direction breakings
starting from 4-D string models with extra gauge symmetries
including $U(1)_A$.

In this paper, we derive formula of soft SUSY breaking scalar masses 
from 4-D string models within a more generic framework.
We consider effects of extra gauge symmetry breakings, 
that is, $D$-term and $F$-term contributions, particle mixing effects
and heavy-light mass mixing effects.
Some phenomenological implications are discussed based on our mass formula.
In particular,
we study the degeneracy and the positivity of squared scalar masses
in special cases.
In addition, we examine specific features of scalar potential
and scalar masses taking an explicit model.

This paper is organized as follows.
In the next section, we explain our starting point reviewing 
the structure of effective SUGRA derived from SST in a field theory
limit.
In section 3, we derive formula of soft SUSY breaking scalar masses
and discuss the phenomenological implications.
In subsection 3.1, flat directions in the SUSY limit are discussed
in the framework of SUGRA.
In subsection 3.2, we discuss classification of scalar fields.
In subsection 3.3, we examine the existence of
heavy-light mass mixing.
In subsection 3.4, a generic formula of soft scalar masses is given.
In subsection 3.5, phenomenological implications are discussed.
In section 4, the results in section 3 are applied 
to an explicit model.
In section 5, we remark some extensions.
Section 6 is devoted to conclusions and discussions.
In Appendix A, formulae of the K\"ahler metric and its inverse
which we use are summarized.

\section{Effective SUGRA as a Field Theory Limit of String Models}

The effective SUGRAs are derived from $Z_N$ orbifold models
taking a field theory limit.
Here we assume the existence of a realistic effective 
SUGRA, that is,
our starting theory has the following excellent features.

The gauge group is 
$G=G'_{SM} \times U(1)^n \times U(1)_A \times H'$ where 
$G'_{SM}$ is a group which contains the standard model gauge group 
$G_{SM} = SU(3)_C \times SU(2)_L \times U(1)_Y$ as a subgroup, $U(1)^n$ are 
non-anomalous $U(1)$ symmetries, $U(1)_A$ is an anomalous $U(1)$
symmetry and $H'$ is a direct product of some non-abelian symmetries.
The anomalies related to $U(1)_A$ are canceled by 
the Green-Schwarz mechanism \cite{GS}.
When gauginos of $H'$ condense, they can trigger SUSY breaking
\cite{gaugino}.
Or $H'$ might be broken by VEVs of some scalar fields at a higher 
energy scale.
We take a standpoint that an origin of SUSY breaking is unspecified. 

Chiral multiplets $\Phi^I$ are classified into two categories.
One is a set of chiral multiplets whose scalar components $\phi^i$
have large VEVs of $O(M)$.
Here $M$ is the gravitational scale defined as 
$M \equiv M_{Pl}/\sqrt{8\pi}$ and $M_{Pl}$ is the Planck scale.
The dilaton field $S$ and the moduli fields $T_{ij}$ belong to $\{\Phi^i\}$.
For the present, we treat only the overall moduli field $T$
($T=T_1=T_2=T_3$, $T_{ij}=0$ for $i \neq j$) and also neglect 
moduli fields $U_i$ corresponding to complex structure.
Further we neglect effects of threshold corrections and an $S$-$T$ mixing.
Later we will discuss the case with several moduli fields $T_i$ and $U_i$
and the case that K\"ahler potential has an $S$-$T$ mixing term.
The other is a set of matter multiplets denoted as $\Phi^{\kappa}$
which contains the MSSM matter multiplets and Higgs multiplets.
Some of them have non-zero $U(1)_A$ ($U(1)^n$, $H'$) charges and 
can induce to the $U(1)_A$ ($U(1)^n$, $H'$) breaking at 
high energy scales by getting VEVs.
We denote the above two types of multiplet as $\Phi^I$ together.
The matter multiplets correspond to massless string states
one to one.

We suppose the following situations related to extra gauge symmetry
breakings.
\begin{enumerate}
\item The $U(1)_A$ symmetry is broken by VEVs of $S$ and
some chiral matter multiplets.

\item Some parts of $U(1)^n$ and $H'$ are broken at much higher energy 
scales than the weak scale by VEVs of some chiral matter multiplets.
Those VEVs are smaller than those of $S$ and $T$, i.e.
\begin{eqnarray}
\langle \phi^{\kappa} \rangle \ll \langle S \rangle, 
\langle T \rangle = O(M) .
\label{<phi>}
\end{eqnarray}
This condition is justified from the fact that a $D$-term condensation
of $U(1)_A$ vanishes up to $O(m_{3/2}^2)$ as will be shown.
Here $m_{3/2}$ is the gravitino mass defined later.

\item The rest extra gauge symmetries are broken spontaneously or 
radiatively by the SUSY breaking effects at some lower scales.
\end{enumerate}
It is straightforward to apply our method to more complicated situations.

We give a comment here.
Such a symmetry breaking generates an intermediate scale $M_I$,
which is defined as the magnitude of VEVs of scalar fields,
below the Planck scale $M_{Pl}$.
Using the ratio $M_I/M_{Pl}$, higher dimensional couplings could explain 
hierarchical structures in particle physics like the fermion masses and 
their mixing angles.
Recently  much attention  has been paid to such a study 
on the fermion mass matrices \cite{texture,texture2}.
In Refs.\cite{texture}, $U(1)$ symmetries are used to generate realistic 
fermion mass matrices and some of them are anomalous, while stringy 
selection rules on nonrenormalizable couplings 
are used in Refs.\cite{texture2}.

Next let us explain the three constituents,
the K\"ahler potential $K$,
the superpotential $W$ and the gauge kinetic function $f_{\alpha\beta}$,
in effective SUGRAs derived from SSTs.
Orbifold models lead to the following K\"ahler potential $K$: 
\cite{ST-SG,OrbSG,OrbSG2}
\begin{eqnarray}
   K &=& -ln(S + S^* + \delta_{GS}^{A} V_A) - 3ln(T + T^*)
\nonumber\\ 
&~&  + \sum_{\kappa} (T + T^*)^{n_\kappa}|\phi^\kappa|^2 + \cdots
\label{K}
\end{eqnarray}
where $\delta_{GS}^{A}$ is a coefficient of the Green-Schwarz
mechanism to cancel the $U(1)_A$ anomaly and 
$V_A$ is a vector superfield of $U(1)_A$.
Here and hereafter we take $M = 1$ according to circumstances.
The dilaton field $S$ transforms nontrivially as 
$S \rightarrow S-i\delta_{GS}^{A}\theta(x)$ under $U(1)_A$ 
with the transformation parameter $\theta(x)$.
The coefficient $\delta_{GS}^{A}$ is given as 
\begin{eqnarray}
  \delta_{GS}^{A} &=& {1 \over 96\pi^2}Tr Q^A 
\label{delta_GS}
\end{eqnarray}
where $Q^A$ is a $U(1)_A$ charge operator.
Further $n_{\kappa}$'s are modular weights of matter fields $\phi^{\kappa}$. 
The formulae of $n_\kappa$ are given in Ref.\cite{OrbSG2,IL}.
The same K\"ahler potential is derived from Calabi-Yau models 
with the large $T$ limit up to twisted sector field's contributions.
If the VEV of $\phi^{\kappa}$ is comparable with one of $T$, 
we should replace 
the second and third terms in Eq.~(\ref{K}) as 
\begin{eqnarray}
-3ln(T + T^*-\sum_{\kappa}|\phi^{\kappa}|^2)
 \end{eqnarray}
for the untwisted sector and  
\begin{eqnarray}
-ln [( T + T^*)^3-\sum_{\kappa}(T + T^*)^{n_{\kappa}+3}|\phi^{\kappa}|^2 ]
\end{eqnarray}
for the twisted sector.

The superpotential $W$ consists of the following two parts,
\begin{eqnarray}
W &=& W_{NP} + W_{Pert} .
\end{eqnarray}
Here $W_{NP}$ is a superpotential induced by some non-perturbative 
effects, and it is expected that VEVs of $S$ and $T$ are 
fixed and SUSY is broken by this part.
The other part $W_{Pert}$ is a superpotential at the tree level 
and starts from trilinear couplings for massless fields
\begin{eqnarray}
W_{Pert} = \sum_{\kappa, \lambda, \mu}
              f_{\kappa\lambda\mu}\phi^{\kappa}\phi^{\lambda}
                                  \phi^{\mu} + \cdots
\end{eqnarray}
where Yukawa couplings $f_{\kappa\lambda\mu}$ generally depend on
the moduli fields $T$ and
the ellipsis stands for terms of higher orders in $\phi^\kappa$.
Note that if the above superpotential includes mass terms 
as $m_{\kappa \lambda}\phi^\kappa \phi^\lambda$, a natural order of 
these masses is of $O(M)$.
Thus we do not include these fields with mass terms at the tree level.
The total K\"ahler potential $G$ is defined as $G \equiv K + ln|W|^2$.
The gauge kinetic function $f_{\alpha\beta}$ is given as
$f_{\alpha\beta} = S \delta_{\alpha\beta}$.
For simplicity, here we assume that Kac-Moody levels satisfy 
$k_\alpha=1$ 
because our results on soft terms are independent of a value of $k_\alpha$.
The scalar potential is given as
\begin{eqnarray}
   V &=& V^{(F)} + V^{(D)} ,
\nonumber\\
   V^{(F)} &\equiv& e^G(G^I (G^{-1})_I^J G_{J}-3),
\label{V(F)}\\
V^{(D)} &\equiv& \frac{1}{2} (Re f^{-1})_{\alpha \beta} D^{\alpha} D^{\beta}
\nonumber\\
      &=& 
{1 \over S+S^*} (K_{\kappa} (T^a \phi)^{\kappa})^2 
+ {1 \over S+S^*} ({\delta_{GS}^{A} \over S+S^*}
  +  K_\kappa (Q^A \phi)^{\kappa})^2 
\nonumber\\
&~&+ {1 \over S+S^*} (K_{\kappa} (Q^B \phi)^{\kappa})^2
 +  {1 \over S+S^*} (K_{\kappa} (T^C \phi)^{\kappa})^2 
\label{V(D)}
\end{eqnarray}
where $G_{I}=\partial G/\partial \phi^I$ and 
$G^{J}=\partial G/\partial \phi_{J}^*$, and
$(Re f^{-1})_{\alpha \beta}$ and  $(G^{-1})_I^{J}$
are the inverse matrices of $Re f_{\alpha \beta}$ and $G_I^{J}$,
respectively.
And the indices $I$, $J$,... run all scalar species,
the index $a$ ($B$, $C$) runs generators of the $G'_{SM}$ 
($U(1)^n$, $H'$) gauge group
and $Q^B$'s are $U(1)^n$ charge operators.  
Note that the Fayet-Iliopoulos $D$-term appears 
in $V^{(D)}$ for $U(1)_A$ if we replace $S$ 
by its VEV \cite{ST-FI,flat1,flat2}.

By the use of K\"ahler potential (\ref{K}),
$D$-terms for $U(1)_A$ and $U(1)^n$ are given as
\begin{eqnarray}
   D^A &=& {\delta_{GS}^{A} \over S+S^*}
  + \sum_\kappa (T+T^*)^{n_\kappa} q_\kappa^A |\phi^\kappa|^2
\label{DA}
\end{eqnarray}
and
\begin{eqnarray}
   D^B &=& \sum_\kappa (T+T^*)^{n_\kappa} q_\kappa^B |\phi^\kappa|^2
\label{DB}
\end{eqnarray}
where $q_\kappa^{A(B)}$ is the $U(1)_A$$(U(1)^n)$ charge of 
scalar field $\phi^\kappa$ 
and we use the relation 
$(Q^{A(B)})^\kappa_\lambda = q^{A(B)}_\kappa \delta^\kappa_\lambda$.

Finally let us give our assumption on the SUSY breaking.
The gravitino mass $m_{3/2}$ is given by
\begin{eqnarray}
   m_{3/2} = \langle e^{K/2M^{2}} {W \over M^2} \rangle
\label{m}
\end{eqnarray}
where $\langle \cdots \rangle$ denotes 
the VEV of the quantity. 
In the next section, it will be often taken to be real as a phase convention.
The $F$-auxiliary fields of the chiral multiplets $\Phi^I$ 
are defined as
\begin{eqnarray}
   F^I \equiv Me^{G/2M^{2}} (G^{-1})_J^{I} G^J.
\label{F}
\end{eqnarray}
It is assumed that SUSY is broken by the $F$-term condensations
of $\phi^i$ such that\footnote{
It is also applicable to the case of SUSY breaking 
by gaugino condensations \cite{gaugino}
because the dynamics are effectively described by a non-perturbative
superpotential for $\phi^i$ after integrating out gauginos.}
\begin{eqnarray}
   \langle F^i \rangle = O(m_{3/2}M).
\label{<F>}
\end{eqnarray}
In this case, stationary conditions of $V$ by $\phi^I$ require 
that VEVs of $D$-auxilary fields should be very small, i.e. 
$\langle D^\alpha \rangle \leq O(m_{3/2}^2)$ and 
$\langle V^{(D)} \rangle$ should vanish up to $O(m_{3/2}^4)$, i.e. 
$\langle V^{(D)} \rangle = O(m_{3/2}^4)$ \cite{JKY,Kawa1}.

\section{Derivation of Soft Scalar Mass Formula}

\subsection{On flat directions}

The effective theories derived from SSTs have, in general,
flat directions in the SUSY limit, which can be a source to 
break gauge symmetries \cite{flat1}.
In this subsection, we discuss such flat directions 
in the framework of SUGRA with $U(1)_A$.
The reasons are as follows.
First we should classify scalar fields 
in a well-defined manner to derive the low-energy effective theory.
That is, we need to specify light fields which appear 
in a low-energy spectrum.
Second there is a possibility that breaking scales of 
extra gauge symmetries can be determined by the existence of $U(1)_A$
and the introduction of SUSY breaking effects.
It is known that some of them are fixed from $D$-flatness condition
of $U(1)_A$ in the SUSY limit \cite{flat1}.
We discuss this possibility from a general viewpoint of scalar
potential in SUGRA.

Let us discuss the classification of scalar fields using mass spectra.
The conditions that SUSY is not spontaneously broken 
in the sector related to matter multiplets are simply expressed as
\begin{eqnarray}
 \frac{\partial \widehat{W}}{\partial \phi^\kappa} &=& 0,
~~~~\widehat{W} \equiv \langle e^{K \over 2M^2} \rangle W,
        \label{SUSYF}\\
{D}^\alpha &=& 0.
        \label{SUSYD}
\end{eqnarray}
Here $\phi^i$'s are replaced by their VEVs in $\widehat{W}$ and 
$D^\alpha$.\footnote{
We assume that VEVs of $\phi^i$ are determined by solving
stationary conditions $\partial V/\partial \phi^i =0$.}
We denote solutions of the above conditions as 
$\phi^{\kappa} = \phi_0^{\kappa}$. 
There, in general, exist several flat directions
and then the magnitudes of $\phi_0^{\kappa}$ are not fixed along such 
flat directions.

In the presence of SUSY breaking,
the vacuum $\langle \phi^I \rangle$ is obtained
by solving the stationary condition 
$\partial V / \partial \phi^I = 0$.

We have the following two kinds of classification 
of scalar fields using $\phi_0^{\kappa}$
or $\langle \phi^\kappa \rangle$ and mass matrices.
\begin{enumerate}
\item In global SUSY models, the supersymmetric fermion mass 
$\mu_{\kappa\lambda}$ is given as
\begin{eqnarray}
\mu_{\kappa\lambda} &=& \frac{\partial^2 \widehat{W}}
{\partial \phi^\kappa \partial \phi^\lambda}|_0 
        \label{mukl}
\end{eqnarray}
where $\cdots|_0$ denotes the value of the quantity 
in the SUSY limit.
By using a basis of $\phi^{\kappa}$ to diagonalize 
$\mu_{\kappa\lambda}$, we can classify scalar fields.

\item In SUGRA, the supersymmetric fermion mass $M_{IJ}$ is given as
\begin{eqnarray}
M_{IJ} &=& \langle Me^{G/2M^2} ( G_{IJ} + {G_{I}G_{J} \over M^2}
 - G_{I'}(G^{-1})^{I'}_{J'} G^{J'}_{IJ} ) \rangle
        \label{MIJ}
\end{eqnarray}
where the VEV is estimated at the minimum $\langle \phi^I \rangle$
of $V$ in the presence of SUSY breaking.
We take a basis of $\phi^I$ to diagonalize the SUSY
fermion mass matrix $M_{IJ}$ and can classify $\phi^I$'s using them.

\end{enumerate}

If we know relations among the above two classifications
in advance, it is enough to use the most convenient one
for our purpose.
Thus let us discuss the relation between $\phi_0^{\kappa}$
and $\langle \phi^\kappa \rangle$.
We estimate the order of $\hat{\delta} \phi^\kappa$ 
where $\langle \phi^\kappa \rangle = \phi_0^{\kappa} + 
\hat{\delta} \phi^\kappa$.
The expansion of $V$ around $\phi_0^{\kappa}$ is given as
\begin{eqnarray}
 V= V|_0 + {\partial V \over \partial \phi^{\kappa}}|_0 
\hat{\delta}\phi^{\kappa} 
+ {\partial^2 V \over \partial \phi^{\kappa}\partial \phi^*_{\lambda}}|_0 
\hat{\delta}\phi^{\kappa}\hat{\delta}\phi^*_{\lambda}
\nonumber\\ 
+ {1 \over 2}{\partial^2 V \over \partial \phi^{\kappa}
\partial \phi^{\lambda}}|_0 
\hat{\delta}\phi^{\kappa}\hat{\delta}\phi^{\lambda} + {\rm H.c.}
+ \cdots .
\end{eqnarray}
Magnitudes of coefficients are estimated as
\begin{eqnarray}
&~&{\partial V \over \partial \phi^{\kappa}}|_0 = O(m_{3/2}\Lambda^2), 
\label{Vkappa0}\\
&~&\frac{\partial ^2 V}{\partial \phi^\kappa \partial \phi^*_\lambda}|_0,
~~~\frac{\partial ^2 V}{\partial \phi^\kappa \partial \phi^\lambda}|_0
= O(\Lambda^2)
\label{Vkappalambda0}
\end{eqnarray}
where we assume that both VEVs of some scalar fields and masses of
heavy fields are of $O(\Lambda)$, that is,
$\phi_0 = O(\Lambda)$, $m_{3/2}G_{KL}|_0 = O(\Lambda)$. 
In string models, this assumption holds for heavy scalar masses
whose origin is Higgs mechanism through VEVs of some scalar fields. From 
the stationary condition 
$\partial V/\partial \langle \phi^\kappa \rangle = 0$,
we find that $\hat{\delta}\phi^\kappa = O(m_{3/2})$.\footnote{
There can be several solutions to satisfy the stationary condition. 
Here we pick out the solution such that
$\langle \phi^\kappa \rangle$ $=$ $\phi_0^{\kappa}$ $+$ $O(m_{3/2})$.}
\footnote{We owe this estimation to the discussion with I.~Joichi and
M.~Yamaguchi.}

In this way, we have the following conclusion.
If there exists a local minimum solution 
$\langle \phi^\kappa \rangle$,
its position is very near to $\phi_0^{\kappa}$,
i.e., 
$\langle \phi^\kappa \rangle$ $=$ $\phi_0^{\kappa}$ $+$ $O(m_{3/2})$. 
By the use of this relation,
the following relations are derived
\begin{eqnarray}
\langle W_{\kappa\lambda} \rangle  &=& W_{\kappa\lambda}|_0 
+ O(m_{3/2}) ,\\
M_{\kappa\lambda} &=& \mu_{\kappa\lambda} + O(m_{3/2}) 
~~~({\rm up~to~phase~factor}) .
\label{M&mu}
\end{eqnarray}
Hence scalar fields are classified into ``heavy'' fields and
``light'' ones using $\mu_{\kappa\lambda}$ in the next subsection.

Let us study SUSY breaking effects on flat directions in effective SUGRAs
derived from string models.
There are several flat directions before SUSY breaking.
After SUSY breaking, the vacuum energy $V_0 \equiv \langle V \rangle$, 
in general, includes VEVs corresponding to flat directions.
Hence such flat directions can be lifted due to SUSY breaking 
and corresponding VEVs can be determined so that a minimum of 
$V_0$ is realized.

The gravitino mass $m_{3/2}$ is written as 
\begin{eqnarray}
m_{3/2}^2 = \langle e^G \rangle 
=\langle e^{K(S,T)}W(S,T)\exp [\sum_\kappa (T+T^*)^{n_\kappa}
|\phi^\kappa|^2] \rangle
\end{eqnarray}
where $K(S,T)$ represents the K\"ahler potential only for $S$ and $T$.
Here we rescale the gravitino mass as 
\begin{eqnarray}
m_{3/2}^2 \rightarrow m_{3/2}^2 e^{f(1)}.
\label{mrescale}
\end{eqnarray}
Here and hereafter $f(a_\kappa)$ denotes 
\begin{eqnarray}
f(a_\kappa) \equiv 
\langle \sum_\kappa a_\kappa (T+T^*)^{n_\kappa}
|\phi^\kappa|^2 \rangle.
\label{fdef}
\end{eqnarray}
In the above Eq.(\ref{mrescale}), $a_\kappa$ is unity.

For a type of K\"ahler potential such as Eq.~(\ref{K}), we can write 
$V_0$ up to $O(m_{3/2}^4)$ as 
\begin{eqnarray}
V_0 & = & V_0(S,T) + V_0^{(M)},\\
V_0(S,T) & \equiv & \langle e^G [(K^S_S)^{-1}|G^S|^2 
\nonumber \\
&~&~~~~~ + (K^T_T)^{-1}|G^T+(K^T_T)(K^{-1})^T_\kappa G^\kappa|^2-3] 
\rangle,
\label{V0ST} \\
V_0^{(M)} &\equiv& 
\langle e^{G} G^\kappa ((G^{-1})^\lambda_\kappa - 
(G^{-1})^T_\kappa(K^T_T)(G^{-1})^\lambda_T) G_\lambda \rangle 
\label{V0(M)}
\end{eqnarray}
where $(K^I_J)$ is a reciprocal of $(K^{-1})^I_J$.
Here from Eqs.~(\ref{m}), (\ref{F}) and (\ref{<F>}),
the magnitude of $V_0(S,T)$ is estimated as $O(m_{3/2}^2 M^2)$. 
Recall that $D$-term contributions to $V_0$ are of $O(m_{3/2}^4)$.
Using the notation of (\ref{fdef}) and the results in Appendix A, 
we can write $V_0^{(M)}$ as 
\begin{eqnarray}
 V_0^{(M)} & = & m_{3/2}^2e^{f(1)}{v^{(M)}_0 \over (T+T^*)^2},
\label{V0f1} \\
 v^{(M)}_0 & = & {f(n^2_\kappa)[9-3f(n_\kappa)-3f(n^2_\kappa)
-f^2(n^2_\kappa)+f(n_\kappa)f(n^2_\kappa)] \over 
(3-f(n_\kappa))^2}
\label{V0f2}
\end{eqnarray}
up to $O(m_{3/2}^4)$.
The vacuum energy $V_0$ depends on VEVs through $f(1)$, $f(n_\kappa)$ 
and $f(n^2_\kappa)$.
Hence SUSY breaking lifts a potential along flat directions 
corresponding VEVs appearing in $f(1)$, $f(n_\kappa)$ and 
$f(n^2_\kappa)$.
Note that when we derive the above results, we use 
the K\"ahler potential (\ref{K}), which is available for 
$\langle \phi^\kappa \rangle \ll \langle T \rangle$.
Thus Eqs.(\ref{V0f1}) and (\ref{V0f2}) are available for 
$f(n_\kappa), f(n^2_\kappa) \ll 1$.
In this limit, we have $ v^{(M)}_0 \approx f(n^2_\kappa)$.

In general, $f(1)$, $f(n_\kappa)$ and $f(n^2_\kappa)$ are 
independent linear combinations of VEVs.
By the definition, we have $f(n^2_\kappa) \geq 0$.
It is obvious that the minimum of $f(n^2_\kappa)$ 
corresponds to the minimum of $v^{(M)}_0$.
At first, we study the case without an anomalous $U(1)$ symmetry.
In this case, $v^{(M)}_0$ as well as $V_0$ takes the minimum 
$v^{(M)}_0=0$ up to $O(m_{3/2}^2)$ at $f(n^2_\kappa)=0$ 
for $f(n^2_\kappa) \ll 1$.
This minimum point means $|\langle \phi^\kappa \rangle|^2 =0$ 
up to $O(m_{3/2}^2)$ for any fields with $n_\kappa \neq 0$, 
because $n_\kappa^2 \geq 0$.
In this case, the vanishing cosmological constant $V_0=0$ requires 
$V_0(S,T)=0$.

In Ref.\cite{flat3}, generic flat directions of $Z_{2n}$ orbifold 
models are discussed.
In these flat directions, pairs of fields with $R$ and $\overline R$ 
representations in the same twisted sector, $n_R=n_{\overline R}$, 
develop their VEVs as 
$\langle R \rangle = \langle \overline R \rangle \neq 0$.
The SUSY breaking effects lift these flat directions determining 
these VEVs $\langle R \rangle = \langle \overline R \rangle =0$ 
up to $O(m_{3/2})$.
Note that these VEVs do not exactly vanish and they could lead 
to symmetry breaking even though these VEVs are very small.

Next we consider the case with an anomalous $U(1)$ symmetry.
In this case, some parts of VEVs are related with 
$\langle \delta_{GS} / (S+S^*) \rangle$ because of the $D$-flatness 
due to the anomalous $U(1)$.
Thus VEVs are written as 
\begin{eqnarray}
\langle (T+T^*)^{n_\kappa}|\phi^\kappa|^2 \rangle = 
\sum_i a_{i \kappa}v_i+a_{X \kappa}\langle {\delta_{GS} \over S+S^*} 
\rangle
\end{eqnarray}
where $v_i$'s are independent degrees of freedom of VEVs and 
$v_i\geq 0$.
Note that some of the coefficients $a_{i\kappa}$, in general, can be 
negative, although $\sum_i a_{i \kappa}v_i+
a_{X \kappa}\langle \delta_{GS}/(S+S^*) \rangle$ should 
satisfy 
\begin{eqnarray}
\sum_i a_{i \kappa}v_i+a_{X \kappa}\langle {\delta_{GS} \over S+S^*} 
\rangle \geq 0.
\label{anomv}
\end{eqnarray}
Here we have 
\begin{eqnarray}
f(n^2_\kappa) = \sum_i \sum_\kappa a_{i \kappa}n_\kappa^2 v_i
+\sum_\kappa a_{X \kappa}n_\kappa^2 \langle {\delta_{GS} \over S+S^*} 
\rangle.
\end{eqnarray}
If $\sum_\kappa a_{i \kappa}n_\kappa^2 >0$, the minimum of 
$f(n^2_\kappa)$ is obtained at $v_i=0$ for a finite value of 
$\langle S \rangle$.
On the other hand, the VEVs $v_i$ corresponding to 
$\sum_\kappa a_{i \kappa}n_\kappa^2 <0$ take the largest values 
at the minimum of $f(n^2_\kappa)$ satisfying Eq.(\ref{anomv}).
Their VEVs $v_i$ are at most of 
$O(\langle \delta_{GS}/(S+S^*)\rangle)$.
Thus values of $f(n^2_\kappa)$ and $v^{(M)}_0$ are of 
$O(\langle \delta_{GS}/(S+S^*)\rangle)$ at the minimum point.
The vanishing cosmological constant $V_0=0$ requires 
$V_0(S,T)=0$ up to $O(\langle \delta_{GS}/(S+S^*)\rangle)$.

In some cases, we have $\sum_\kappa a_{i \kappa}n_\kappa^2=0$, and 
either of $f(n_\kappa^2)$ and $v^{(M)}_0$ does not include 
corresponding VEVs $v_i$.
Thus the directions along $v_i$ with 
$\sum_\kappa a_{i \kappa}n_\kappa^2=0$ are still flat directions 
of $V_0$ up to $O(m_{3/2}^4)$.
As will be shown, soft scalar masses depend on $f(q_\kappa)$, 
$f(q_\kappa n_\kappa)$ and $f(q_\kappa^2)$.
If the VEVs $v_i$ with $\sum_\kappa a_{i \kappa}n_\kappa^2=0$ 
appear in these functions, $f(q_\kappa)$, $f(q_\kappa n_\kappa)$ 
and $f(q_\kappa^2)$, a potential along flat directions of 
$v_i$ are lifted at the level of $O(m_{3/2}^4)$.
In models with anomalous $U(1)$ symmetries, $D$-term 
contributions on soft masses are dominantly obtained by 
VEVs including $\langle \delta_{GS}/(S+S^*)\rangle$ and 
$v_i$ with $\sum_\kappa a_{i \kappa}n_\kappa^2<0$, because 
these VEVs are of $O(\langle \delta_{GS}/(S+S^*)\rangle)$ 
and other VEVs $v_i$ with 
$\sum_\kappa a_{i \kappa}n_\kappa^2>0$ are of 
$O(m_{3/2}^2)$.

In this way, a breaking scenario we supposed in section 2 can be realized.
Then the symmetry breakings at a very large scale are induced by 
$D$-flatness of $U(1)_A$ and the order is given as
$O(\langle \delta_{GS}/(S+S^*)\rangle^{1/2})$.
We denote it by $M_I$ and it is estimated as 
$O(10^{-1} M) \sim O(10^{-3} M)$ by using explicit models.
Other symmetry breakings can occur by the SUSY breaking effects spontaneously
or radiatively at $O(m_{3/2})$.

\subsection{Classification of scalar fields}

We take the basis $\hat{\phi}^\kappa$ 
that the supersymmetric fermion mass $\mu_{\kappa\lambda}$ 
is diagonalized in the SUSY limit.
Then we assume that scalar fields are classified 
into ``heavy'' fields $\hat{\phi}^K, \hat{\phi}^L, \cdots$ and  
``light'' fields $\hat{\phi}^k, \hat{\phi}^l, \cdots$, 
such as $\mu_{KL}=O(M_I)$ and $\mu_{kl}=O(m_{3/2})$, respectively.
In string models, all light fields are massless, i.e. $\mu_{kl}=0$,
in the SUSY limit.
The diagonalized fields $\hat{\phi}^\kappa$ are given 
as linear combinations of original ones ${\phi}^\kappa$ (string states)
such as
\begin{eqnarray}
\hat{\phi}^\kappa = R^\kappa_\lambda \phi^\lambda .
        \label{R}
\end{eqnarray}

The K\"ahler potential of matter parts $K^{(M)}$ is, in general, 
written as
\begin{eqnarray}
K^{(M)} &=& K_\kappa^\lambda \phi^\kappa \phi^*_\lambda
               + H_{\kappa\lambda} \phi^\kappa \phi^\lambda + {\rm H.c.}
\nonumber\\
&=& \hat{K}_\kappa^\lambda \hat{\phi}^\kappa \hat{\phi}^*_\lambda
               + \hat{H}_{\kappa\lambda} \hat{\phi}^\kappa \hat{\phi}^\lambda
          + {\rm H.c.} ,\\
\hat{K}_\kappa^\lambda &\equiv& (R^{-1})_\kappa^{\mu}
          K_{\mu}^{\nu} R_{\nu}^\lambda ,\\
\hat{H}_{\kappa\lambda} &\equiv& H_{\mu\nu}
          (R^{-1})_\kappa^{\mu} (R^{-1})_\lambda^{\nu} 
\label{K(M)}
\end{eqnarray}
where the second (third) term in RHS is
a holomorphic (anti-holomorphic) part on matter fields.
The $D$-terms are also written as
\begin{eqnarray}
D^{\alpha} &=& K_\kappa (T^{\alpha})^\kappa_\lambda \phi^\lambda
= \hat{K}_\kappa (\hat{T}^{\alpha})^\kappa_\lambda \hat{\phi}^\lambda ,\\
\hat{K}_\kappa &\equiv& (R^{-1})_\kappa^{\mu} K_{\mu} ,\\
(\hat{T}^{\alpha})^{\kappa}_{\lambda} &\equiv& R^\kappa_{\mu}
 (T^{\alpha})^{\mu}_{\nu}(R^{-1})_{\lambda}^{\nu} 
\label{hatD}
\end{eqnarray}
for a linear-realization of gauge symmetries.

Next we discuss the case of the effective SUGRA derived from 
$Z_N$ orbifold models.
The K\"ahler potential of matter parts $K^{(M)}$ is given as
\begin{eqnarray}
K^{(M)} &=& \sum_{\kappa,\lambda,\mu} 
(R^{-1})_{\kappa}^\mu (T+T^*)^{n_\mu} R^\lambda_{\mu}
 \hat{\phi}^{\kappa} \hat{\phi}^*_{\lambda} 
\label{K(M)string}
\end{eqnarray}
where we neglect the contribution of moduli fields $U_i$.
$D$-terms for $U(1)$ symmetries are given as
\begin{eqnarray}
   D^A &=& {\delta_{GS}^{A} \over S+S^*}
  + \sum_{\kappa,\lambda,\mu} 
(R^{-1})_{\kappa}^\mu (T+T^*)^{n_\mu} q_\mu^A R^\lambda_{\mu}
 \hat{\phi}^{\kappa} \hat{\phi}^*_{\lambda} ,
\label{hatDA}\\
   D^B &=& 
   \sum_{\kappa,\lambda,\mu} 
(R^{-1})_{\kappa}^\mu (T+T^*)^{n_\mu} q_\mu^B R^\lambda_{\mu}
 \hat{\phi}^{\kappa} \hat{\phi}^*_{\lambda} .
\label{hatDB}
\end{eqnarray}
As discussed in section 2, Yukawa couplings has a moduli-dependence,
so $\mu_{\kappa\lambda}$ and $R^\kappa_\lambda$, in general, depend
on the VEV of the moduli field $T$.

\subsection{Heavy-light mixing terms}

In this subsection, we estimate magnitudes of heavy-light
mixing mass terms 
$\langle V_H^k \rangle \equiv \langle \partial^2 V/\partial \hat{\phi}^H
\partial \hat{\phi}^*_k \rangle$
and $\langle V_{Hk} \rangle \equiv \langle \partial^2 V/\partial \hat{\phi}^H
\partial \hat{\phi}^k \rangle$
including both the SUSY part and the soft SUSY breaking part.
We use the vacuum $\langle \hat{\phi}^{\kappa} \rangle$ in place of 
$\hat{\phi}^\kappa_0$ since the difference between the estimation by
$\langle \hat{\phi}^{\kappa} \rangle$ and that by $\hat{\phi}^\kappa_0$
is the quantity of $O(m_{3/2}^2)$.
After some calculations, $\langle V_H^k \rangle$ is
expressed as
\begin{eqnarray}
\langle V_H^k \rangle &=& M_{HH'} 
\langle (\hat{K}^{-1})^{H'}_l \rangle M^{lk}
\nonumber\\
&~&- M_{HH'} \langle (\hat{K}^{-1})^{H'}_{J} \rangle 
  \langle (\hat{K})^{Jk}_I \rangle \langle \hat{F}^I \rangle
+ O(m_{3/2}^2)
\label{VH^k}
\end{eqnarray}
where $(\hat{K}^{-1})^I_J$ is the inverse matrix of 
$(\hat{K})^I_J$.
Carets represent functions of $\hat{\phi}^I$ and $\hat{\phi}^*_J$.
If there are heavy-light mixing terms of O(1) in
the K\"ahler potential, the order of the first term 
in the RHS of Eq.(\ref{VH^k}) can be $O(m_{3/2} M_I)$.
This contribution is SUSY one and is negligible except the stop and
Higgs doublets
in the MSSM since $M_{kl}$ is very small compared with the weak scale.
The second term in the RHS of Eq.(\ref{VH^k}) can be of $O(m_{3/2} M_I)$
if the K\"ahler potential has heavy-light mixing terms of $O(1)$ and 
a holomorphic part.

The chirality flipped scalar mass $\langle V_{Hk} \rangle$ is
written as
\begin{eqnarray}
\langle V_{Hk} \rangle &=&
 m_{3/2} \langle \hat{G}_{HkI} \rangle \langle \hat{F}^I \rangle
\nonumber\\
&~& - M_{HH'} \langle (\hat{K}^{-1})^{H'}_{J} \rangle 
  \langle (\hat{K})^{J}_{kI} \rangle \langle \hat{F}^I \rangle
+ O(m_{3/2}^2) .
\label{VHk}
\end{eqnarray}
If there are Yukawa couplings of $O(1)$ among heavy, light and moduli
fields in the superpotential,
the order of the first term in the RHS of Eq.(\ref{VHk})
can be $O(m_{3/2} M)$.
The second term in the RHS of Eq.(\ref{VHk}) can be of $O(m_{3/2} M_I)$
if the K\"ahler potential has heavy-light mixing terms of $O(1)$.
If there are Yukawa couplings of $O(1)$ among light and moduli
fields in the superpotential, that is,
$m_{3/2} \langle \hat{G}_{klI} \rangle = O(1)$, 
the order of the scalar masses for light fields
can be $O(m_{3/2} M)$ and the weak scale can be destabilized in the
presence of weak scale Higgs doublets with such intermediate masses.
This is so called $\lq\lq$ gauge hierarchy problem".
Only when $m_{3/2}\langle \hat{G}_{IJJ'} \rangle
 \langle \hat{F}^{J'} \rangle$'s meet
some requirements, the hierarchy survives.
In many cases, we require the following condition,
\begin{eqnarray}
m_{3/2}\langle \hat{G}_{IJJ'} \rangle \langle \hat{F}^{J'}
\rangle \leq
O(m_{3/2}^2)
\label{gh1}
\end{eqnarray}
for light fields $\hat{\phi}^I$ and $\hat{\phi}^J$ \cite{JKY,Kawa1}.
If we impose the same condition to the light fields 
and the heavy fields,
we neglect the effect of the first term in the RHS of Eq. (\ref{VHk}).
In such a case, unless there exist heavy-light mixing terms
in the K\"ahler potential, there appear no heavy-light mass mixing terms
of $O(m_{3/2} M_I)$.
In string models, whether there exist heavy-light mixing terms
in the K\"ahler potential or not is model-dependent.

\subsection{Soft scalar masses}

\subsubsection{Soft scalar mass terms}

For convenience, we introduce new notations related to
the classification of scalar fields,
\begin{eqnarray}
 \hat{\phi}^{\cal H} \equiv 
\left(
\begin{array}{c}
\Delta \hat{\phi}^{H} \\
\Delta \hat{\phi}^*_{H}
\end{array}
\right) 
=
\left(
\begin{array}{c}
\hat{\phi}^{H} - \langle \hat{\phi}^{H} \rangle \\
\hat{\phi}^*_{H} - \langle \hat{\phi}^*_{H} \rangle
\end{array}
\right) ,~~~
\hat{\phi}^{\cal L} \equiv 
\left(
\begin{array}{c}
\hat{\phi}^{l} \\
\hat{\phi}^*_{l}
\end{array}
\right)
\label{phical}
\end{eqnarray}
where $\hat{\phi}^{H}$'s are heavy scalar fields
and $\hat{\phi}^{l}$'s are light scalar fields.
We assume that there are no light fields whose VEVs are of $O(M_I)$.
The light fields can get VEVs of $O(m_{3/2})$ and induce extra
gauge symmetry breakings, but we treat them 
as a sum of the VEVs and flactuations since both have a same order and
our goal is to derive soft scalar mass formula at $M_I$.
We can take effects of symmetry breakings at $O(m_{3/2})$
in the same way.

Scalar mass terms are written as
\begin{eqnarray}
 V^{\rm Mass}&=&\frac{1}{2} \{ \hat{\phi}^*_{\cal H} 
{\sl H}^{\cal H}_{\cal H'} \hat{\phi}^{\cal H'}
+ \hat{\phi}^*_{\cal H} {\sl M}^{\cal H}_{\cal L'} \hat{\phi}^{\cal L'}
\nonumber\\
&~&~~~~ + \hat{\phi}^*_{\cal L} {\sl M}^{\dagger{\cal L}}_{\cal H'}
 \hat{\phi}^{\cal H'}
+ \hat{\phi}^*_{\cal L} {\sl L}^{\cal L}_{\cal L'} \hat{\phi}^{\cal L'} \}
\label{Vmass}
\end{eqnarray}
where 
\begin{eqnarray}
 {\sl H} &\equiv& {\sl H}^{\cal H}_{\cal H'} =
\left(
\begin{array}{cc}
\langle V^H_{H'} \rangle & \langle V^{HH'} \rangle \\
\langle V_{HH'} \rangle & \langle V_{H}^{H'} \rangle 
\end{array}
\right) ,
\label{slH}\\
 {\sl M} &\equiv& {\sl M}^{\cal H}_{\cal L'} =
\left(
\begin{array}{cc}
\langle V^H_{l'} \rangle & \langle V^{Hl'} \rangle \\
\langle V_{Hl'} \rangle & \langle V_{H}^{l'} \rangle 
\end{array}
\right) ,
\label{slM}\\
 {\sl L} &\equiv& {\sl L}^{\cal L}_{\cal L'} =
\left(
\begin{array}{cc}
\langle V^l_{l'} \rangle & \langle V^{ll'} \rangle \\
\langle V_{ll'} \rangle & \langle V_{l}^{l'} \rangle 
\end{array}
\right) .
\label{slL}
\end{eqnarray}
The order of the above mass matrices are estimated as
${\sl H} = O(M_I^2)$, ${\sl M} = O(m_{3/2} M_I)$
and ${\sl L} = O(m_{3/2}^2)$.
Scalar mass terms are rewritten as
\begin{eqnarray}
 V^{\rm Mass}=\frac{1}{2} ( \hat{\phi}^*_{\cal H} + \hat{\phi}^*_{\cal L}
{\sl M}^{\dagger} {\sl H}^{-1} ) {\sl H}
( \hat{\phi}^{\cal H'} + {\sl H}^{-1} {\sl M}\hat{\phi}^{\cal L'})
\nonumber\\
-\frac{1}{2} \hat{\phi}^*_{\cal L} {\sl M}^{\dagger} {\sl H}^{-1} 
 {\sl M}\hat{\phi}^{\cal L'}
+\frac{1}{2} \hat{\phi}^*_{\cal L} {\sl L} \hat{\phi}^{\cal L'} .
\label{Vmass'}
\end{eqnarray}
We discuss implication of each term in Eq.(\ref{Vmass'}).
The first term is the mass term among heavy fields.
After the integration of the heavy fields
$\hat{\phi}^{\cal H} + {\sl H}^{-1} {\sl M}\hat{\phi}^{\cal L}$,
there appear $D$-term \cite{D-term,KMY2,Kawa1} and extra $F$-term 
contributions \cite{KMY2,Kawa1} to scalar
masses which will be discussed later.
The second term is new contributions which appear 
after the diagonalization of scalar mass terms.
This contribution can be sizable, i.e., $O(m_{3/2}^2)$, if the heavy-light
mass mixing is $O(m_{3/2} M_I)$.
The last term is a mass term among light fields.
Note that the heavy fields 
$\hat{\phi}^{\cal H} + {\sl H}^{-1} {\sl M}\hat{\phi}^{\cal L}$
and the light fields $\hat{\phi}^{\cal L}$ used here are different from 
properly diagonalized fields up to $O(m_{3/2}/M_I)$ terms.
The final expressions of scalar masses of $O(m_{3/2}^2)$
for light fields are same whichever we use 
as a definition of the scalar fields.
We give a more fully expression for extra contribution 
due to the exiatence of heavy-light mass mixing terms as 
\begin{eqnarray}
 V_{\rm Soft~Mass}^{\rm Mix} &\equiv& 
-\frac{1}{2} \hat{\phi}^*_{\cal L} {\sl M}^{\dagger} {\sl H}^{-1} 
 {\sl M}\hat{\phi}^{\cal L'}
\nonumber\\
&=& (V_{\rm Soft~Mass}^{\rm Mix})_l^k \hat{\phi}^l \hat{\phi}^*_k
+ {1 \over 2}(V_{\rm Soft~Mass}^{\rm Mix})_{kl} \hat{\phi}^k \hat{\phi}^l +
{\rm H.c.} 
\label{VMix}
\end{eqnarray}
where
\begin{eqnarray}
(V_{\rm Soft~Mass}^{\rm Mix})_l^k &=&
- \langle V_{lH} (V^{-1})_{H'}^{H} V^{H'k} \rangle
- \langle V_{lH} (V^{-1})^{HH'} V_{H'}^{k} \rangle
\nonumber\\
&~& - \langle V_{l}^{H} (V^{-1})_{HH'} V^{H'k} \rangle
- \langle V_{l}^{H} (V^{-1})_{H}^{H'} V_{H'}^{k} \rangle ,
\label{DeltaVMix1}\\
(V_{\rm Soft~Mass}^{\rm Mix})_{kl} &=&
- \langle V_{kH} (V^{-1})_{H'}^{H} V^{H'}_{l} \rangle
- \langle V_{kH} (V^{-1})^{HH'} V_{H'l} \rangle
\nonumber\\
&~& - \langle V_{k}^{H} (V^{-1})_{HH'} V^{H'}_{l} \rangle
- \langle V_{k}^{H} (V^{-1})_{H}^{H'} V_{H'l} \rangle .
\label{DeltaVMix2}
\end{eqnarray}

\subsubsection{Parametrization}

For the analysis of soft SUSY breaking parameters, 
it is convenient to introduce the following parameterization 
\begin{eqnarray}
&~&\langle e^{G/2} (K^S_S)^{-1/2} G^S \rangle 
 = \sqrt{3} C m_{3/2} e^{i\alpha_S} \sin\theta ,
\label{G^S}\\
&~&\langle e^{G/2} (K^T_T)^{-1/2} (G^T + 
(K^T_T)(K^{-1})^T_\kappa G^\kappa)
 \rangle  = \sqrt{3} C m_{3/2} e^{i\alpha_T} \cos\theta 
\label{G^T}
\end{eqnarray}
where $(K^I_J)$ is a reciprocal of $(K^{-1})^I_J$.
Using Eq.~(\ref{V0ST}), the vacuum energy $V_0$ is written as
\begin{eqnarray}
V_0 &=& \langle e^G(G^I (G^{-1})_I^J G_{J}-3) \rangle
\nonumber\\
 &=& 3 (C^2 - 1) m_{3/2}^2 + V_0^{(M)} 
\label{V0-para}
\end{eqnarray}
up to $O(m_{3/2}^4)$.
Since $C^2$ should be positive or zero,
we have a constraint $V_0^{(M)} \leq 3m_{3/2}^2 +V_0$ from 
Eq.~(\ref{V0-para}).
In the case with $V_0=0$, it becomes as
\begin{eqnarray}
\langle G^\kappa ((G^{-1})^\lambda_\kappa - 
(G^{-1})^T_\kappa (K^T_T) (G^{-1})^\lambda_T) G_\lambda \rangle \leq 3.
\label{Con}
\end{eqnarray}
It gives a constraint on VEVs of $\phi^\kappa$ and $T$.
Further a larger value of $V_0^{(M)}$ in the above region means 
$C \ll 1$.
Such a limit as $C \to 0$ corresponds to the \lq\lq moduli-dominated" 
breaking, that is, $\langle F^S \rangle \ll 1$ and 
$\langle F^T \rangle$ and $\langle F^\kappa \rangle$ contribute 
to the SUSY-breaking.
Note that this situation does not agree with the case of 
the moduli-dominated 
breaking without extra gauge symmetry breakings $\sin \theta \to 0$.
The relation (\ref{<phi>}) 
implies the relation
$\langle G^\kappa \rangle \ll \langle G^S \rangle , \langle G^T \rangle$
because we discuss the vacuum solution
near to the flat direction which leads to the relations
$\langle W_\kappa \rangle \equiv 
\langle \partial W/\partial \phi^\kappa \rangle = O(m_{3/2}^2)$ and
$\langle G_\kappa \rangle = \langle K_\kappa \rangle + O(m_{3/2})$.
In this case, the parameterization (\ref{G^T}) becomes 
a simpler one such as
\begin{eqnarray}
&~&\langle e^{G/2} (K^T_T)^{-1/2} G^T  \rangle  
= \sqrt{3} C m_{3/2} e^{i\alpha_T} \cos\theta .
\label{G^T'}
\end{eqnarray}
Further the discussion in 3.1 means that $V_0$ is dominated by 
$V_0(S,T)$ because of $V_0(S,T) \gg V_0^{(M)}$.
Thus the vacuum energy $V_0$ becomes as
\begin{eqnarray}
V_0 &=& 3 (C^2 - 1) m_{3/2}^2 .
\label{V0-para'}
\end{eqnarray}

Soft SUSY breaking scalar mass terms are given as 
\begin{eqnarray}
V_{\rm Soft~Mass}^{(0)} &=& (m_{3/2}^2 + V_0) \sum_{\kappa}
\langle (T+T^*)^{n_\kappa} \rangle|\phi^{\kappa}|^2 
\nonumber\\
&+& \sum_{\kappa} \langle F^I \rangle \langle K_{I\kappa}^{I'}
(K^{-1})^{J'}_{I'} K_{J'}^{J\kappa} - K_{I\kappa}^{J\kappa} \rangle
\langle F^*_J \rangle
\langle (T+T^*)^{n_\kappa} \rangle |\phi^{\kappa}|^2
\label{SoftMass}
\end{eqnarray}
before heavy fields are integrated out.
By the use of the parametrization and the diagonalized fields
$\hat{\phi}^\kappa$,
$V_{\rm Soft~Mass}^{(0)}$ is rewritten as 
\begin{eqnarray}
V_{\rm Soft~Mass}^{(0)} &=& \sum_{\kappa,\lambda} (m_{3/2}^2 + V_0) 
\langle \hat{K}_{\kappa}^\lambda \rangle \hat{\phi}^{\kappa} 
\hat{\phi}^*_\lambda 
 + \sum_{\kappa,\lambda} m_{3/2}^2 C^2 \cos^2\theta 
\hat{N}_{\kappa}^\lambda \hat{\phi}^{\kappa} \hat{\phi}^*_\lambda
\label{SoftMass'}
\end{eqnarray}
where
\begin{eqnarray}
\hat{K}_\kappa^\lambda &\equiv& (R^{-1})_\kappa^{\mu}
          (T+T^*)^{n_{\mu}} R_{\mu}^\lambda ,\\
\hat{N}_\kappa^\lambda &\equiv& 
\langle \hat{K}_\kappa^\mu \rangle \hat{n}_\mu^\lambda ,
~~~~\hat{n}_\kappa^\lambda \equiv (R^{-1})_\kappa^{\nu}
          n_{\nu} R_{\nu}^\lambda .
\end{eqnarray}
After heavy fields are integrated out,
we have the following mass terms for light fields at the energy scale $M_I$,
\begin{eqnarray}
V_{\rm Soft~Mass} &=& \sum_{k,l} (m_{3/2}^2 + V_0) \langle \hat{K}_{k}^{l} 
\rangle \hat{\phi}^{k} \hat{\phi}^*_{l} 
+ \sum_{k,l} m_{3/2}^2 C^2 \cos^2\theta \hat{N}_{k}^l
\hat{\phi}^{k} \hat{\phi}^*_l 
\nonumber\\
&+& V_{\rm Soft~Mass}^{\rm D} + V_{\rm Soft~Mass}^{\rm Extra~F}
+ V_{\rm Soft~Mass}^{\rm Mix} + V_{\rm Soft~Mass}^{\rm Ren}
\label{SoftMass''}
\end{eqnarray}
where $V_{\rm Soft~Mass}^{\rm D}$, $V_{\rm Soft~Mass}^{\rm Extra~F}$,
$V_{\rm Soft~Mass}^{\rm Mix}$ and $V_{\rm Soft~Mass}^{\rm Ren}$
are $D$-term contributions, extra $F$-term
contributions which will be discussed in the following subsections,
the contributions due to the existence of heavy-light mass mixing
discussed in the previous section and contributions of renormalization
effects from $M$ to $M_I$, respectively.

\subsubsection{$D$-term contributions}

The $D$-term contributions are given as \cite{D-term,KMY2,Kawa1},
\begin{eqnarray}
V_{\rm Soft~Mass}^{\rm D} &=& \sum_{k,l}
\sum_\alpha g^2_\alpha \langle D^\alpha \rangle 
(\hat{Q}^\alpha)_k^l \hat{\phi}^k \hat{\phi}^*_l
\label{SoftMassD}
\end{eqnarray}
where
\begin{eqnarray}
(\hat{Q}^\alpha)_k^l &\equiv& \langle \hat{K}_k^\mu \rangle 
(\hat{q}^\alpha)_\mu^l ,~~~~
(\hat{q}^\alpha)_\kappa^\lambda \equiv (R^{-1})_\kappa^{\nu}
          q^\alpha_{\nu} R_{\nu}^\lambda .
\end{eqnarray}
Here $g_\alpha$'s are gauge coupling constants
and we use the relation $\langle ReS \rangle = 1/g_{\alpha}^2$.
We omit the terms whose magnitudes are less than $O(m_{3/2}^4)$.

Next we rewrite $V_{\rm Soft~Mass}^{\rm D}$ using the parametrization 
introduced before.
For this purpose, it is useful to adapt to the following relation 
of $D$-term condensations \cite{Kawa1},
\begin{eqnarray}
g_\alpha \langle D^\alpha \rangle =
2 (M_V^{-2})^{\alpha\beta} g_\beta \langle F^I \rangle
\langle F^*_J \rangle \langle (D^\beta)_I^J \rangle 
\label{D-cond}
\end{eqnarray}
where $(M_V^{-2})^{\alpha\beta}$ is the inverse matrix of
gauge boson mass matrix $(M_V^2)^{\alpha\beta}$ given as
\begin{eqnarray}
(M_V^2)^{\alpha\beta} = 
  2 g_{\alpha} g_{\beta} \langle (T^\beta(\phi^\dagger))_I
K^I_J (T^\alpha(\phi))^J \rangle .
        \label{MV2}
\end{eqnarray}
Here the gauge transformation of $\phi^I$ is 
given as $\delta \phi^I = i g_{\alpha} (T^\alpha (\phi))^I$
up to space-time dependent infinitesimal parameters.

After some straightforward calculations,
$D$-term condensations are written as
\begin{eqnarray}
g^2_{\hat{\alpha}} \langle D^{\hat{\alpha}} \rangle &=& 2 
g_{\hat{\alpha}} m_{3/2}^2 
\{ (M_V^{-2})^{\hat{\alpha} A} g_A (1 - 6C^2 \sin^2\theta )
\langle \sum_\kappa {q_\kappa^A} (T+T^*)^{n_\kappa} |\phi^\kappa|^2 \rangle
\nonumber \\
&~& ~~~~ -  \sum_{\hat{\beta}} (M_V^{-2})^{\hat{\alpha}\hat{\beta}}
g_{\hat{\beta}} C^2 \cos^2\theta 
\langle \sum_\kappa q_\kappa^{\hat{\beta}} n_\kappa (T+T^*)^{n_\kappa}
 |\phi^\kappa|^2 \rangle \} 
\label{D-cond2}
\end{eqnarray}
where $\hat{\alpha}$ and $\hat{\beta}$ run over broken generators.

We give some comments.
We need to introduce three kinds of model-dependent quantities such as
\begin{eqnarray}
&~&\langle \sum_\kappa q_\kappa^A (T+T^*)^{n_\kappa} |\phi^\kappa|^2 \rangle ,
\\
&~&\langle \sum_\kappa q_\kappa^{\hat{\alpha}} q_\kappa^{\hat{\beta}}
 (T+T^*)^{n_\kappa} |\phi^\kappa|^2 \rangle ,
\\
&~&\langle \sum_\kappa q_\kappa^{\hat{\alpha}} n_\kappa (T+T^*)^{n_\kappa} 
|\phi^\kappa|^2 \rangle .
\end{eqnarray}
Their magnitudes are of order $O(M_I^2)$.
Note that 
$\langle \sum_\kappa q_\kappa^{\hat{\alpha}'} (T+T^*)^{n_\kappa} 
|\phi^\kappa|^2 \rangle = O(m_{3/2}^2)$ due to the $D$-flatness
condition in the presence of SUSY breaking.
Here $\hat{\alpha}'$ runs over only non-anomalous diagonal broken generators.

In the case that there are no mixing elements between $U(1)_A$ and
other symmetries in $(M_V^2)^{\alpha\beta}$, the mass of $U(1)_A$ 
gauge boson is given as
\begin{eqnarray}
(M_V^2)^A &=&   2 g_A^2 \{ \langle \sum_\kappa q_\kappa^A 
(T+T^*)^{n_\kappa} |\phi^\kappa|^2 \rangle^2
\nonumber\\
&~&~~~~ + \langle \sum_\kappa (q_\kappa^A)^2 (T+T^*)^{n_\kappa}
|\phi^\kappa|^2 \rangle \}
        \label{MA2}
\end{eqnarray}
by the use of the relation $Q^A S = \delta_{GS}$ and 
$(Q^A \phi)^\lambda = q_\lambda^A \phi^\lambda$.
Under the assumption that $\langle \phi^\kappa \rangle \ll M$,
$(M_V^2)^A$ is simplified as
\begin{eqnarray}
(M_V^2)^A =   2 g_A^2 \langle \sum_\kappa (q_\kappa^A)^2 
(T+T^*)^{n_\kappa} |\phi^\kappa|^2 \rangle 
        \label{MA2'}
\end{eqnarray}
and $D$-term condensation of $U(1)_A$ is written as
\begin{eqnarray}
g^2_A \langle D^A \rangle &=& {m_{3/2}^2 \over 
\langle \sum_\kappa (q_\kappa^A)^2 (T+T^*)^{n_\kappa} 
|\phi^\kappa|^2 \rangle} \times 
\nonumber\\
&~&\{(1 - 6C^2 \sin^2\theta )
\langle \sum_\kappa {q_\kappa^A} (T+T^*)^{n_\kappa} |\phi^\kappa|^2 \rangle
\nonumber\\
&~&~~~~ - C^2 \cos^2\theta 
\langle \sum_\kappa {q_\kappa^A} n_\kappa (T+T^*)^{n_\kappa}
 |\phi^\kappa|^2 \rangle \} .
\label{DA-cond}
\end{eqnarray}
Furthermore, at that time, $D$-term condensations of non-anomalous
symmetries are given as
\begin{eqnarray}
g^2_{\hat{\alpha}'} \langle D^{\hat{\alpha}'} \rangle &=&
 - m_{3/2}^2 C^2 \cos^2\theta 
{\langle \sum_\kappa {q_\kappa^{\hat{\alpha}'}} n_\kappa (T+T^*)^{n_\kappa}
 |\phi^\kappa|^2 \rangle  \over
\langle \sum_\kappa (q_\kappa^{\hat{\alpha}'})^2 
(T+T^*)^{n_\kappa} |\phi^\kappa|^2 \rangle} 
\label{DB-cond}
\end{eqnarray}
where broken charges are re-defined by the use of diagonalization of 
$(M_V^2)^{\hat{\alpha}'\hat{\beta}'}$.
Using the expression (\ref{DB-cond}),
we can show that there appears no sizable $D$-term contribution 
to scalar masses
if a broken symmetry is non-anomalous and SUSY is broken by the dilaton
$F$-term, i.e., $\cos^2\theta = 0$.

In a simple case that only one field $X$, which has no charges except 
the $U(1)_A$ charge, gets VEV to cancel the contribution
of $S$ in $D^A$, the above result is reduced to the previous one 
obtained in Ref.\cite{U1A}.
\footnote{In the formula obtained in Ref.\cite{U1A}, 
there is a sign error:
$+6C^2\sin^2\theta$ should be $-6C^2\sin^2\theta$.}
Note that our result is not reduced 
to that obtained from the theory with the Fayet-Iliopoulos $D$-term, 
which is derived from the effective SUGRA by taking the flat 
limit first \cite{N}, 
even in the limit that $|\delta_{GS}^A/q_X^A| \ll 1$.
This disagreement originates from the fact that we regard 
$S$ and $T$ as dynamical fields, that is, we use the stationary conditions 
$\partial V/ \partial \phi^I = 0$ to calculate $D$-term condensations.

The dominant $D$-term contributions to mixing mass terms
are obtained as
\begin{eqnarray}
\sum_\alpha g^2_\alpha \langle D^\alpha \rangle 
\langle K^\lambda_{kl} \rangle \langle \phi^*_\lambda \rangle 
{\phi}^k (T^\alpha {\phi})^l + {\rm H.c.} .
\label{SoftMassDmix}
\end{eqnarray}
The magnitude of them are estimated as $O(m_{3/2}^4 M_I/M)$
and so they are neglected in the case that $M_I \ll M$. 
Note that the contribution of $O(m_{3/2}^4)$ such as
$g^2_\alpha \langle D^\alpha \rangle \langle H_{\kappa\lambda} \rangle 
\phi^\kappa (T^\alpha \phi)^\lambda$ vanishes from gauge invariance of 
the holomorphic part $H$ of the K\"ahler potential.

\subsubsection{Extra $F$-term contributions}

After the integration of complex heavy fields $\hat{\phi}^K$
and Nambu-Goldstone multiplets $\hat{\phi}^{\hat{\alpha}}$, 
the following $F$-term contributions appear
in the low-energy effective scalar potential \cite{KMY2,Kawa1},
\begin{eqnarray}
V^{\rm Extra~F} &=& V_{(0)}^{\rm Extra~F} + V_{(1)}^{\rm Extra~F}
 + V_{(2)}^{\rm Extra~F}
\label{VextraF}\\
V_{(0)}^{\rm Extra~F} &\equiv& 
            E\{- \delta^2 ({\hat{W}^* \over M^2}\hat{{\cal G}}^K)
\langle (\hat{K}^{-1})_K^L \rangle 
                 \delta^2 ({\hat{W} \over M^2}\hat{{\cal G}}_L)
\nonumber \\
&~&~~~ 
+ \delta^2 ({\hat{W}^* \over M^2}\hat{{\cal G}}^{\hat{\alpha}})
\langle (\hat{K}^{-1})_{\hat{\alpha}}^{\hat{\beta}} \rangle 
            \delta^2 ({\hat{W} \over M^2}\hat{{\cal G}}_{\hat{\beta}})
\nonumber \\
&~&~~~ 
+ \delta ({\hat{W}^* \over M^2}\hat{{\cal G}}^{\hat{\alpha}})
\langle (\hat{K}^{-1})_{\hat{\alpha}}^{\hat{\beta}} \rangle 
            \delta^{3'}({\hat{W} \over M^2}\hat{{\cal G}}_{\hat{\beta}})
 + {\rm H.c.}
\nonumber \\
&~&~~~ + \delta^2 ({\hat{W}^* \over M^2}\hat{{\cal G}}^K)
\langle (\hat{K}^{-1})_K^L \rangle 
 \biggl(\langle \hat{W}_{Lk} \rangle \hat{\phi}^k +
{1 \over 2}\langle \hat{W}_{Lkl} \rangle \hat{\phi}^{k} \hat{\phi}^{l} 
+ \cdots \biggr)
\nonumber \\
&~&~~~ + {\rm H.c.} \} ,
\label{V(F)0}\\
V_{(1)}^{\rm Extra~F} &\equiv& 
            E\{
\delta ({\hat{W}^* \over M^2}\hat{{\cal G}}^{K})
\langle (\hat{K}^{-1})_{K}^{\hat{\beta}} \rangle 
            \delta^{3'}({\hat{W} \over M^2}\hat{{\cal G}}_{\hat{\beta}})
 + {\rm H.c.}
\nonumber \\
&~&~~~ + \delta ({\hat{W}^* \over M^2}\hat{{\cal G}}^\kappa)
\langle \delta (\hat{K}^{-1})_\kappa^L \rangle 
          \delta^{2'} ({\hat{W} \over M^2}\hat{{\cal G}}_L )
  + {\rm H.c.} 
\nonumber \\
&~&~~~
+ \delta ({\hat{W}^* \over M^2}\hat{{\cal G}}^\kappa)
\langle \delta^{2'} (\hat{K}^{-1})_\kappa^\lambda \rangle 
              \delta ({\hat{W} \over M^2}\hat{{\cal G}}_\lambda)\} ,
\label{V(F)1}\\
V_{(2)}^{\rm Extra~F} &\equiv& 
            E\{
\delta^{2} ({\hat{W}^* \over M^2}\hat{{\cal G}}^{\hat{\alpha}})
\langle (\hat{K}^{-1})_{\hat{\alpha}}^k \rangle 
            \delta^{2}({\hat{W} \over M^2}\hat{{\cal G}}_{k})
 + {\rm H.c.}
\nonumber \\
&~&~~~ + \delta ({\hat{W}^* \over M^2}\hat{{\cal G}}^\kappa)
\langle \delta (\hat{K}^{-1})_\kappa^k \rangle 
          \delta^{2} ({\hat{W} \over M^2}\hat{{\cal G}}_k)
  + {\rm H.c.} \} .
\label{V(F)2}
\end{eqnarray}
We explain our notations in the above Eqs.(\ref{V(F)0})--(\ref{V(F)2}).
Carets represent functions of $\hat{\phi}^I$ and $\hat{\phi}^*_J$.
The quantity $\hat{{\cal G}}_\lambda$ is defined as
\begin{eqnarray}
\hat{{\cal G}}_\lambda \equiv \hat{G}_\lambda + 
            (\hat{G})_\lambda^\kappa (\hat{G}^{-1})_\kappa^j \hat{G}_j
\end{eqnarray}
and $E$ is defined as $E \equiv \langle {\rm exp}(\hat{K}/M^2) \rangle$.
Any quantity $A$ is expanded in powers of $m_{3/2}$ such as
\begin{eqnarray}
A = \langle A \rangle + \delta A + \delta^2 A + \cdots .
\label{expU}
\end{eqnarray}
For example, 
\begin{eqnarray}
\delta ({\hat{W} \over M^2} \hat{\cal G}_K) &=& 
\langle \hat{W}_K \rangle +
 \langle \hat{W}_{KL} \rangle \delta \hat{\phi}^L
+ {\langle \hat{W} \rangle \over M^2}\langle \hat{K}_K \rangle 
\nonumber \\
&+& \langle (\hat{K})_{K}^{\lambda} \rangle 
  \langle (\hat{K}^{-1})_\lambda^{j} \rangle 
  \delta ({\hat{W} \over M^2} \hat{G}_j),
\label{gK1}
\\
\delta^2 ({\hat{W} \over M^2} \hat{\cal G}_K) &=& 
\langle \hat{W}_{KL} \rangle \delta^2 \hat{\phi}^L
+ \langle \hat{W}_{Kk} \rangle \delta \hat{\phi}^k
+ \langle \hat{W}_{K\hat{\beta}} \rangle 
\delta \hat{\phi}^{\hat{\beta}}
+ {1 \over 2}\langle \hat{W}_{K\lambda\mu} \rangle 
\delta \hat{\phi}^{\lambda} \delta \hat{\phi}^{\mu} 
\nonumber \\
&+& {1 \over M^2}\biggl({1 \over 2}\langle \hat{W}_{LM} \rangle 
\delta \hat{\phi}^L \delta \hat{\phi}^M \langle \hat{K}_K \rangle
+ \langle \hat{W} \rangle \langle \hat{K}_{KJ} \rangle 
\delta \hat{\phi}^J 
 \nonumber \\
&+& \langle \hat{W} \rangle \langle \hat{K}_{K}^{J} \rangle 
\delta \hat{\phi}_J \biggr) \\
&+& \langle (\hat{K})_{K}^\lambda \rangle 
\langle (\hat{K}^{-1})_{\lambda}^j \rangle
  \delta^2 ({\hat{W} \over M^2} \hat{G}_j)
+ \delta ((\hat{K})_{K}^\lambda (\hat{K}^{-1})_{\lambda}^j)
  \delta ({\hat{W} \over M^2} \hat{G}_j) .
\nonumber
\label{gK2}
\end{eqnarray}
When there exist flat directions,
we have $\langle \hat{W}_{\lambda} \rangle = O(m_{3/2}^2)$.
Quantities with a prime such as 
$\delta^{3'}({\hat{W} \over M^2}\hat{{\cal G}}_{\hat{\beta}})$
mean that the terms proportional to 
$\delta^{2} \hat{\phi}^{I}$ are omitted.
The ellipsis in Eq.~(\ref{V(F)0}) represents other terms 
in $\delta^{2}({\hat{W} \over M^2}\hat{{\cal G}}_K)
 - \langle \hat{W}_{KL} \rangle \delta^2 \hat{\phi}^L$.

The first and second terms in (\ref{V(F)1}) are estimated as 
$O(m_{3/2}^4 M_I/M)$
and the last one is $O(m_{3/2}^4 (M_I/M)^2)$.
They are negligible in the case that $M_I \ll M$.
Scalar masses of $O(m_{3/2}^2)$
are given as
\begin{eqnarray}
V_{\rm Soft~Mass}^{\rm Extra~F} &=& (V_{\rm Soft~Mass}^{\rm Extra~F})_k^l
\hat{\phi}^k \hat{\phi}^*_l ,
\label{VSoftextraF}\\
(V_{\rm Soft~Mass}^{\rm Extra~F})_k^l &\equiv& 
  E\{- \langle (\hat{W}^{-1})^{LM} \rangle
 \delta ({\hat{W}^* \over M^2}\hat{{\cal G}}^\kappa)
\langle (\hat{K}^{-1})_\kappa^{\hat{\alpha}} \rangle 
\langle \hat{W}_{\hat{\alpha}Mk} \rangle 
\langle \hat{K}_L^K \rangle 
\nonumber\\
&~&~~~~\times \langle \hat{W}^{*{\hat{\beta}Nl}} \rangle 
\langle (\hat{K}^{-1})_{\hat{\beta}}^{\lambda} \rangle 
 \delta ({\hat{W} \over M^2}\hat{{\cal G}}_\lambda)
\langle (\hat{W}^{*-1})_{NK} \rangle
\nonumber\\
&~&~~~+ \biggl(\langle \hat{W}_{\hat{\alpha}k} \rangle
+ {\langle \hat{W} \rangle \over M^2} 
\langle (\hat{K})_{\hat{\alpha}k} \rangle \biggr)
\langle (\hat{K}^{-1})^{\hat{\alpha}}_{\hat{\beta}} \rangle 
\nonumber\\
&~&~~~~\times \biggl(\langle \hat{W}^{*\hat{\beta}l} \rangle
+ {\langle \hat{W}^* \rangle \over M^2} 
\langle (\hat{K})^{\hat{\beta}l} \rangle \biggr)
\\
&~&~~~+ {\langle \hat{W} \rangle \over M^2} 
\langle (\hat{K})_{\hat{\alpha}}^l \rangle
\langle (\hat{K}^{-1})^{\hat{\alpha}}_{\hat{\beta}} \rangle 
{\langle \hat{W}^* \rangle \over M^2} 
\langle (\hat{K})_k^{\hat{\beta}} \rangle
\nonumber\\
&~&~~~- \langle (\hat{W}^{-1})^{LM} \rangle
 \delta ({\hat{W}^* \over M^2}\hat{{\cal G}}^\kappa)
\langle (\hat{K}^{-1})_\kappa^{\hat{\alpha}} \rangle 
\langle \hat{W}_{\hat{\alpha}Mk} \rangle 
{\langle \hat{W} \rangle \over M^2} 
\langle \hat{K}_L^l \rangle 
\nonumber\\
&~&~~~- {\langle \hat{W}^* \rangle \over M^2} 
\langle \hat{K}_k^K \rangle 
\langle \hat{W}^{*{\hat{\beta}Nl}} \rangle 
\langle (\hat{K}^{-1})_{\hat{\beta}}^{\lambda} \rangle 
 \delta ({\hat{W} \over M^2}\hat{{\cal G}}_\lambda)
\langle (\hat{W}^{*-1})_{NK} \rangle \}.
\nonumber 
\label{MassExtraF}
\end{eqnarray}
We discuss conditions that $(V_{\rm Soft~Mass}^{\rm Extra~F})_k^l$
is neglected.
If Yukawa couplings among Nambu-Goldstone, heavy and light fields
are small enough, the first term and the last two terms are neglected.
If we impose $R$-parity conservation,
the second and third terms in $(V_{\rm Soft~Mass}^{\rm Extra~F})_k^l$ 
are forbidden since bilinear couplings between Nambu-Goldstone  
and light fields are $R$-parity odd.
Here we define the $R$-parity of $\hat{\phi}^\kappa$ as follows,
$$R(\hat{\phi}^{\hat{\alpha}}) = +1,~~~R(\hat{\phi}^{K}) = 
R(\hat{\phi}^{k}) = -1 .$$

\subsubsection{Formula of soft scalar masses}

Using scalar mass terms (\ref{SoftMass''}) and diagonalizing
of the K\"ahler potential, i.e., 
$\langle \hat{K}_k^l \rangle = \delta_k^l$,
we have the following mass formula for light scalar fields 
at the energy scale $M_I$,
\begin{eqnarray}
(m^2)_k^l|_{M_I} &=& (m_{3/2}^2 + V_0) \delta_k^l
+ m_{3/2}^2 C^2 cos^2\theta \hat{N}_{k}^l
\nonumber\\
&+& \sum_{\hat{\alpha}} g^2_{\hat{\alpha}} \langle D^{\hat{\alpha}} \rangle 
(\hat{Q}^{\hat{\alpha}})_k^l 
+ (V_{\rm Soft~Mass}^{\rm Extra~F})_k^l
+ (V_{\rm Soft~Mass}^{\rm Mix})_k^l 
\nonumber\\
&+& (V_{\rm Soft~Mass}^{\rm Ren})_k^l 
\label{SoftMassFormula}
\end{eqnarray}
where $(V_{\rm Soft~Mass}^{\rm Ren})_k^l$ is a sum of contributions 
related to renormalization effects from $M$ to $M_I$ and consists
of the following two parts.
One is a radiative correction between $M$ and $M_I$.
This contribution $(\Delta m^2)_\kappa^\lambda|_{M_I}$ is given as 
\cite{RGE},
\begin{eqnarray}
    (\Delta m^2)_\kappa^\lambda|_{M_I}  & =& 
          - \sum_{\alpha} \frac{2}{b_{\alpha}}
        C_2 (R_\kappa^\alpha)({M_\alpha^2(M_I)} - {M_\alpha^2(M)})
           \delta_\kappa^\lambda
\nonumber\\
        & &   + \sum_B \frac{1}{b_B} Q_{R_\kappa}^{(B)}
                        ({S_{B}(M_I)} - {S_{B}(M)}) 
           \delta_\kappa^\lambda ,
\label{Deltam2} \\
    S_{B}(M_I) &=& \frac{\alpha_B (M_I)}{\alpha_B (M)} 
                       S_{B}(M) ,\\
     S_{B}(\mu) &\equiv& \sum_{R_\kappa} Q_{R_\kappa}^{(B)} 
         n_{R_\kappa} (m^2)_\kappa^\kappa(\mu)
\label{SB} 
\end{eqnarray}
where $\alpha$ runs all the gauge groups 
but $B$ runs only non-anomalous $U(1)$ gauge groups
whose charges are $Q_{R_\kappa}^{(\alpha)}$,
 $C_2 (R_\kappa^\alpha)$'s are the second order Casimir invariants,
 $M_{\alpha}$'s are gaugino masses and
 $n_{R_\kappa}$ is the multiplicity.
Here we neglect effects of Yukawa couplings.
It is straightforward to generalize our results to 
the case with large Yukawa couplings.
Here we use the anomaly cancellation condition 
$\sum_{R_\kappa} C_2 (R_\kappa^\alpha) Q_{R_\kappa}^{(B)} 
n_{R_\kappa} = 0$
and the relation of orthgonality 
$\sum_{R_\kappa} Q_{R_\kappa}^{(B)} Q_{R_\kappa}^{(B')} 
n_{R_\kappa} = b_B \delta_{BB'}$.
Note that there is no contribution related to $U(1)_A$ symmetry 
since it is broken at $M$.

The other is $D$-term contribution due to mass splitting
which is induced by mass renormalization.
We denote it by $(\Delta m_D^2)_k^l|_{M_I}$.
This contribution is given as
\begin{eqnarray}
&~&(\Delta m_D^2)_k^l|_{M_I} = \sum_{\hat{\alpha}} g^2_{\hat{\alpha}} 
\langle \Delta D^{\hat{\alpha}} \rangle (\hat{Q}^{\hat{\alpha}})_k^l 
\end{eqnarray}
where
\begin{eqnarray}
&~&g_{\hat{\alpha}} \langle \Delta D^{\hat{\alpha}} \rangle \equiv
   -2 (M_V^{-2})^{\hat{\alpha}\hat{\beta}} g_{\hat{\beta}}
\langle \sum_{\kappa,\lambda,\mu} \hat{\phi}^*_\lambda 
(\Delta m^2)_\mu^\lambda|_{M_I}
(\hat{Q}^{\hat{\beta}})_\kappa^\mu \hat{\phi}^{\kappa} \rangle .
\label{Deltam2D}
\end{eqnarray}

\subsection{Phenomenological implications of soft scalar mass}

Here we discuss phenomenological implications of
our soft scalar mass formula, especially $D$-term 
contributions, considering simple cases.
In general, $D$-term contributions are comparable with 
$F$-term contributions.
Our formula could lead to 
a strong non-universality of soft scalar masses.
Recently much work is devoted to phenomenological implications of 
the non-universality\cite{KMY1,KMY2,nonuni}.
In addition, various researches of soft scalar masses have been done
in the presence of anomalous $U(1)$ symmetry
\cite{N,U1A,U1X,CD}.
We examine the universality, the degeneracy and the positivity
of squared soft scalar masses in the case that 
there are neither particle mixing in the K\"ahler potential,
heavy-light mass mixing effects nor extra $F$-term contributions.
Here we take $V_0 = 0$, i.e. $C^2=1$ 
and consider the case that there are no mixing elements between 
$U(1)_A$ and other symmetries in $(M_V^2)^{\alpha\beta}$ for simplicity.

\subsubsection{Anomaly-free symmetry case}

Here we consider models with an anomaly-free symmetry.
In this case, soft scalar masses are obtained as 
\begin{eqnarray}
m^2_k = m^2_{3/2}\left[ 1+\cos^2 \theta\left( n_k -q_k 
{\langle \sum_\lambda q_\lambda^{\hat{\alpha}'}n_\lambda (T+T^*)^{n_\lambda} 
|\phi^\lambda|^2 \rangle 
\over 
\langle \sum_\lambda (q_\lambda^{\hat{\alpha}'})^2 (T+T^*)^{n_\lambda} 
|\phi^\lambda|^2 \rangle }\right) \right].
\end{eqnarray}
If $q_k\langle \sum_\lambda q_\lambda^{\hat{\alpha}'}n_\lambda 
(T+T^*)^{n_\lambda} 
|\phi^\lambda|^2 \rangle >0$, squared soft masses $m^2_k$ can 
easily become negative, especially for larger value of 
$\cos \theta$.

Experiments for the process of flavor changing neutral current (FCNC)
 require that $\Delta m^2/m_{3/2}^2  \siml 10^{-2}$ 
for the first and the second families in the case with 
$m^2_{\tilde q} \sim O(1)$TeV\cite{FCNC}.
Hence we should derive $\Delta m^2/m_{3/2}^2  \approx 0$ 
within the level of $O(10^{-2})$.
Hereafter $a \approx 0$ denotes such a meaning.

In the limit that $\cos^2 \theta \rightarrow 0$, i.e. the dilaton 
dominant SUSY breaking, we obtain universal soft scalar masses, 
$m^2_k=m^2_{3/2}$ \cite{Sdomi,BIM}.
In order to realize degenerate soft scalar masses in the other values 
of $\cos \theta$, one needs a ``fine-tuning'' condition for 
differences of $n_k$ and $q_k$, 
$\Delta n$ and $\Delta q$, as 
\begin{eqnarray}
\Delta n =\Delta q
{\langle \sum_\lambda q_\lambda^{\hat{\alpha}'}n_\lambda (T+T^*)^{n_\lambda} 
|\phi^\lambda|^2 \rangle 
\over 
\langle \sum_\lambda (q_\lambda^{\hat{\alpha}'})^2 (T+T^*)^{n_\lambda} 
|\phi^\lambda|^2 \rangle }.
\end{eqnarray}

If fields with non-vanishing VEVs have the same modular weight, 
i.e. the same K\"ahler metric,  
the $D$-term contributions on soft scalar masses vanish due to the 
$D$-flatness condition.
This fact is important.
This situation can happen in some cases.
In these cases, degeneracy of soft masses is realized 
for fields with the same values of modular weights.
One example for the vanishing $D$-term contribution 
is shown in the next section.

Another interesting example is the case where enhanced gauge 
symmetries break by VEVs of moduli fields in orbifold models.
Gauge symmetries are enhanced at specific points of moduli spaces, 
where some massless states $\eta_i$ also appear in the untwisted sector.
For example, $Z_3$ orbifold models have enhanced $U(1)^6$ symmetries.
Here we expand moduli fields $\tau_i$ around these 
points so that vanishing or nonvanishing VEVs $\langle \tau_i \rangle$ 
correspond to unbroken or broken enhanced symmetries.
Neither $\tau_i$ nor $\eta_i$ has well-defined charges under 
the $U(1)$'s and we take linear combinations $s_i$, which have 
definite $U(1)$ charges \cite{enhance}.
These fields $s_i$ have the same K\"ahler metric.
If only these fields $s_i$ develop VEVs and no symmetry other than 
enhanced symmetries break, $D$-term contributions on soft scalar masses 
vanish.
Because enhanced symmetries are anomaly-free and fields developing VEVs 
have the same K\"ahler metric.
These models will be studied in detail elsewhere.

\subsubsection{Anomalous $U(1)$ case}

Here we study models with an anomalous $U(1)$ symmetry.
In this case, soft scalar masses are obtained as \footnote{
Throughout this subsection, we omit the superscript $A$ in the 
$U(1)_A$ charge $q_k^A$.}
\begin{eqnarray}
m^2_k = m_{3/2}^2[1+n_k\cos^2\theta
+{q_k \over f(q^2_\lambda)}\{f(q_\lambda)(1-6\sin^2\theta)
-f(q_\lambda n_\lambda)\cos^2\theta \}],
\end{eqnarray}
where $f(a_\lambda)$ denotes Eq.(\ref{fdef}).
Here $f(q_\lambda)$ does not vanish for a finite value of 
$\langle S \rangle$ because of the $D$-flatness of $U(1)_A$.
It is remarkable that even if $\cos \theta=0$, the $D$-term 
contribution does not vanish.
That is different from $D$-term contributions due to 
the breakdown of anomaly-free $U(1)$ symmetries.
In general, non-universal soft scalar masses are obtained even if 
$\cos \theta =0$.

The $D$-term contribution vanishes if the following fine-tuning 
condition is satisfied 
\begin{eqnarray}
\left( 6f(q_\lambda)-f(q_\lambda n_\lambda)\right)\sin^2 \theta 
=f(q_\lambda)-f(q_\lambda n_\lambda).
\label{D0}
\end{eqnarray}
We have $f(q_\lambda n_\lambda)=n_\lambda f(q_\lambda)$ in the case 
where the fields in the summation $f(q_\lambda n_\lambda)$ have 
the same modular weight $n_\lambda$.
In this case Eq.~(\ref{D0}) reduces as 
\begin{eqnarray}
(6-n_\lambda)\sin^2 \theta = 1-n_\lambda.
\label{ftc}
\end{eqnarray}
For $n_\lambda =1$, the moduli dominant SUSY breaking, i.e.,
$\sin \theta=0$, 
satisfies this condition although this modular weight $n_\lambda =1$ 
is not naturally obtained \cite{IL,mod-wei}.
The modular weight satisfying $n_\lambda \leq 0$ leads to 
$0 < \sin^2 \theta < 1$ for Eq.~(\ref{ftc}).

Degenerate soft scalar masses are obtained for differences of 
modular weights and $U(1)_A$ charges, $\Delta n$ and $\Delta q$, 
in the case where the following fine-tuning condition is satisfied 
\begin{eqnarray}
\left(  f(q^2_\lambda)\Delta n + 6f(q_\lambda) \Delta q 
-f(q_\lambda n_\lambda) \right)\cos^2 \theta = 5f(q_\lambda)\Delta q.
\end{eqnarray}
Soft scalar masses are written for two extreme cases of the 
SUSY breaking, i.e. $\cos \theta=0$ and 1 as 
\begin{eqnarray}
&~& {m^2_k \over m_{3/2}^2}  = 
1-5q_k{ f(q_\lambda) \over f(q^2_\lambda)} \quad  {\rm for} \quad 
\cos \theta=0,\\
&~& {m^2_k \over m_{3/2}^2}   = 
1+n_k+q_k{f(q_\lambda)-f(q_\lambda n_\lambda) \over f(q^2_\lambda)} 
 \quad {\rm for} \quad \cos \theta=1.
\end{eqnarray}
Squared soft scalar masses become easily negative for 
$q_kf(q_\lambda) >0$ in the former case.
On the other hand, we obtain likely negative $m^2_k$ for 
$q_k\{ f(q_\lambda)-f(q_\lambda n_\lambda)\} < 0$ in the latter case.

As a concrete example, we discuss a simple model where only one 
field $X$ develops its VEV.
Its modular weight and anomalous $U(1)$ 
charge are denoted as $n_X$ and $q_X$.
In this case, the soft scalar mass is given as
\begin{eqnarray}
m^2_{k} = m_{3/2}^2[1+n_{k} \cos^2 \theta
 +{q_{k} \over q_X}((6-n_X)\cos^2 \theta -5)] .
\label{mass2}
\end{eqnarray}
Note that the coefficient of $q_k/q_X$ 
in Eq.~(\ref{mass2}) is sizable.

We obtain the difference of the soft masses as 
\begin{eqnarray}
{\Delta m^2 \over m_{3/2}^2}=\Delta n\cos^2 \theta 
+{\Delta q \over q_X}((6-n_X)\cos^2 \theta - 5) 
\end{eqnarray}
by using Eq.~(\ref{mass2}).
If $\Delta q/q_X \approx 0$, we have 
$\Delta m^2 /m_{3/2}^2=\Delta n \cos^2 \theta$.
In this case, the limit $\cos^2 \theta \rightarrow 0$ leads to 
$\Delta m \rightarrow 0$.
It corresponds to the dilaton-dominated breaking, 
where soft masses are universal \cite{Sdomi,BIM}.
Unless $\Delta q/q_X \approx 0$, 
we needs \lq \lq fine-tuning'' on the value of $\cos \theta$ 
as 
\begin{eqnarray}
\cos^2 \theta \approx {5 \over 6-n_X+q_X\Delta n/ \Delta q}.
\label{cos^2}
\end{eqnarray}
This \lq \lq fine-tuning'' is possible only in the case where
\begin{eqnarray}
n_X \leq 1 + {\Delta n \over \Delta q}q_X.
\label{Con2}
\end{eqnarray}
If Eq.~(\ref{cos^2}) is satisfied, 
the soft scalar mass is written as 
\begin{eqnarray}
m_k^2=m_{3/2}^2[1+{5(n_k-\Delta nq_k/\Delta q) \over 
6-n_X+\Delta n q_X / \Delta q}].
\end{eqnarray}

The condition for the positivity of $m^2_{k}$ is written as 
\begin{eqnarray}
-n_k+{q_k \over q_X}(n_X-6) \leq 
(1-5{q_k \over q_X}) \cos^{-2} \theta .
\label{Con-3}
\end{eqnarray}
If $1-5q_k /q_X$ is positive, 
we can find a solution $\cos \theta$ of the above 
constraint for any $n_k, n_X, q_k$ and $q_X$.
On the other hand, if $1-5q_k /q_X$ is negative, 
it leads to the following constraint:
\begin{eqnarray}
1+n_k \geq {q_k \over q_X}(n_X-1),
\end{eqnarray}
because $\cos^{-2} \theta \geq 1$.

Let us consider two extreme examples for the SUSY-breaking, i.e. 
$\cos \theta =0$ and 1.
We have 
\begin{eqnarray}
&~& m_k^2=m_{3/2}^2(1-5{q_k \over q_X}) 
 \quad {\rm for} \quad \cos \theta=0,
\label{Sdom} \\
&~& m_k^2=m_{3/2}^2[1+n_k+{q_k \over q_X}(1-n_X)] 
 \quad {\rm for} \quad \cos \theta=1.
\label{Tdom}
\end{eqnarray}
Matter fields usually have modular weights $n_k \leq 0$ \cite{IL,mod-wei}.
Thus the fields with $q_k/q_X > 0$ for Eq.~(\ref{Sdom}) and 
$q_k/q_X < 0$ for  Eq.~(\ref{Tdom}) can easily have negative 
squared scalar mass of $O(m_{3/2}^2)$ at the Planck scale.
That implies that several fields could develop VEV's and 
they could trigger symmetry breakings.
We can show that there exist fields with $q_k/q_X<0$ 
for each gauge group other than $U(1)_A$.
The reason is as follows.
Let us assume the gauge group is $U(1)_A \times \prod_\ell G_\ell$.
The Green-Schwarz anomaly cancellation mechanism requires that 
$C_{G_\ell}=\delta_{GS}^A k_{\ell}$ for any $\ell$, where $C_{G_\ell}$ is 
a coefficient of $U(1)_A \times G_\ell^2$ anomaly 
and $k_{\ell}$ is a Kac-Moody level of $G_\ell$.
Through the $U(1)_A$ breaking due to the Fayet-Iliopoulos $D$-term, 
the field $X$ develops its VEV. 
Here its charge should satisfy $q_X Tr Q^A < 0$ and $q_X C_{G_\ell} < 0$ 
to satisfy the $D$-flatness of the anomalous $U(1)$.
Each gauge group $G_\ell$ always has fields $\phi^\kappa$
which corresponds nontrivial representation on its group 
and whose $U(1)_A$ charges satisfy $q_X q_k < 0$ 
because of $q_X C_{G_\ell} < 0$.
The $D$-term contribution on soft terms is very sizable.
That could naturally lead to $m_k^2 < 0$ except a narrow region 
and cause  $G_\ell$ breaking.

\section{Analysis on Explicit Model}

\subsection{Flat direction}

In this section, we study $U(1)_A$ breaking effects on flat directions
and derive specific scalar mass relations
by using an explicit model \cite{LRmodel}.
The model we study is the $Z_3$ orbifold model with 
a shift vector $V$ and Wilson lines $a_1$ and $a_3$ such as
$$V={1 \over 3}(1,1,1,1,2,0,0,0)(2,0,0,0,0,0,0,0)' ,$$
$$a_1={1 \over 3}(0,0,0,0,0,0,0,2)(0,0,1,1,0,0,0,0)' ,$$
$$a_3={1 \over 3}(1,1,1,2,1,1,1,0)(1,1,0,0,0,0,0,0)' .$$
This model has a gauge group as 
$$ G = SU(3)_C \times SU(2)_L \times SU(2)_R \times U(1)^7 
\times SO(8)' \times SU(2)'.$$
One of $U(1)^7$ is anomalous.
This model has matter multiplets as 
\begin{eqnarray*}
{\rm U-sec.} &: \quad &3[(3,2,1)_0+(\bar 3,1,2)_0+(1,2,2)_0] \\
&~&+3[(8,2)'_6+(1,1)'_{-12}],
\end{eqnarray*}
\begin{eqnarray*}
{\rm T-sec.} &: \quad &9[(3,1,1)_4+(\bar 3,1,1)_4] 
+15[(1,2,1)_4+(1,1,2)_4]\\
(N_{OSC}=0)&~& +3(1,2,2)_4+
3[(1,2,1)(1,2)'_{-2}+(1,1,2)(1,2)'_{-2}]\\
&~&24(1,2)'_{-2}+60(1,1,1)_4+3(1,1,1)_{-8},
\end{eqnarray*}
$${\rm T-sec.} (N_{OSC}=-1/3):\quad 9(1,1,1)_4$$
where the number of suffix denotes the anomalous $U(1)$ charge 
defined as $Q^A \equiv Q_5 - Q_6$ 
and $N_{OSC}$ is the oscillator number.
This model has $Tr Q^A=864$.
The $U(1)$ charge generators of $U(1)^7$ are defined in Table 1.

This model has many $SU(3)_C \times SU(2)_L \times SU(2)_R$-singlets 
as shown above.
These fields are important for flat directions leading to realistic vacua.
For example, this model includes the following 
$SU(3)_C \times SU(2)_L \times SU(2)_R$-singlets 
\begin{eqnarray}
& u: \ & Q_a= (0,0,0,0,-6,6,0),\nonumber \\
& Y: \ & Q_a= (0,-4,-4,0,2,-2,0),\nonumber \\
& S_1: \ & Q_a= (0,-4,-4,0,-4,4,0),\nonumber \\
& D'_3: \ & Q_a= (6,4,0,-2,-2,0,-2),\nonumber \\
& D'_4: \ & Q_a= (6,4,0,2,-2,0,2),\nonumber \\
& D'_5: \ & Q_a= (-6,0,4,-2,0,2,-2),\nonumber \\
& D'_6: \ & Q_a= (-6,0,4,2,0,2,2)\nonumber 
\end{eqnarray}
where $U(1)^7$ charges $Q_a$ $(a=1,2,\cdots,7)$ are represented 
in the basis of Table 1.
Here we follow the notation of the fields in Ref.~\cite{LRmodel} 
except the $D'_i$ fields.
These $D'_i$ fields are $SU(2)'$-doublets in the non-oscillated 
twisted sector with $n_k=-2$, corresponding to $T_i$ fields in 
Ref.~\cite{LRmodel}.
The others are singlets under any non-abelian group.
The $u$ fields corresponds to the untwisted sector with $n_u=-1$.
In addition, $S_1$ corresponds to the non-oscillated twisted sector 
with $n_k=-2$ and $Y$ corresponds to the twisted sector with a 
nonvanishing oscillator number.
Thus the field $Y$ has the modular weight $n_Y=-3$.
There exist the following flat directions \cite{LRmodel}
\begin{eqnarray}
&~& \langle (T+T^*)^{-1}|u|^2 \rangle = v_1, \\
&~& \langle (T+T^*)^{-3}|Y|^2 \rangle = 
\langle (T+T^*)^{-2}|D'_3|^2 \rangle =
\langle (T+T^*)^{-2}|D'_6|^2 \rangle =v_2, \nonumber \\
&~& \langle (T+T^*)^{-2}|S_1|^2 \rangle = 
\langle (T+T^*)^{-2}|D'_4|^2 \rangle =
\langle (T+T^*)^{-2}|D'_5|^2 \rangle =v_3
\nonumber
\end{eqnarray}
where $v_i \geq 0$.
Along this flat direction, the gauge symmetries break 
as $U(1)^7 \times SU(2)' \rightarrow U(1)^3$.
One of unbroken $U(1)^3$ charges corresponds to $Q_{B-L}$.
Here we define the broken charges as follows,
\begin{eqnarray*}
&~& Q'_1 \equiv {1 \over 3}(2Q_1 + Q_2 - Q_3 - Q_5 - Q_6) ,\\
&~& Q'_2 \equiv {1 \over \sqrt{3}}(2Q_4 + Q_7) ,\\
&~& Q'_3 \equiv {1 \over \sqrt{2}}(Q_2 + Q_3) ,\\
&~& Q^A \equiv  Q_5 - Q_6 ,~~~T'^3
\end{eqnarray*}
where $T'^3$ is a third component of generators of $SU(2)'$.
Note that the gauge boson mass matrix is not diagonalized
in this definition of charges.
The modular weights and broken charges of the light scalar fields 
and the fields with VEVs are given in Table 2.
For the light fields, we follow the notation of fields 
in Ref.\cite{LRmodel}.
Chiral multiplets are denoted as $Q_L$ for left-handed quarks,
$Q_R$ right-handed quarks, $L$ for left-handed leptons 
and $R$ for right-handed leptons.
In addition, $H$ are Higgs doublets.

The $D$-flatness condition for $U(1)_A$ requires 
\begin{eqnarray}
\langle {\delta_{GS}^A \over S+S^*} \rangle - 12v_1-12v_3=0.
\label{GSVV}
\end{eqnarray}
On the top of that, we have 
\begin{eqnarray}
f(n^2_\lambda)=v_1+17v_2+12v_3.
\end{eqnarray}
Under the constraint (\ref{GSVV}), the minimum of $f(n^2_\lambda)$ 
is obtained at the following point:
\begin{eqnarray}
v_1={1 \over 12}\langle {\delta_{GS}^A \over S+S^*} \rangle, \quad 
v_2 \sim O(m_{3/2}^2), \quad  v_3 \sim O(m_{3/2}^2).
\label{vi}
\end{eqnarray}
Using $Tr Q^A=864$ and $\langle ReS \rangle \sim 2$, 
we estimate $v_1 \sim M^2/53$. From Eqs.(\ref{vi}), 
the breaking scale of $U(1)'_i$'s $(i=1,2,3)$ and $SU(2)'$
is estimated as $O(m_{3/2})$.

\subsection{Soft mass relations}

We derive specific relations among soft scalar masses. 
The basic idea and the strategy are the same as those 
in Ref.\cite{KMY1,KMY2,KT}. 
The SUSY spectrum at the weak scale, which is expected to 
be measured in the near future, is translated into the soft 
SUSY breaking parameters.
The values of these parameters at higher energy scales 
are obtained by using the renormalization group 
equations (RGEs) \cite{RGE}.
In many cases, there exist some relations among these parameters. 
They reflect the structure of high-energy physics.
Hence we can specify the high-energy physics by checking
these relations.

The generic formula of scalar mass is given as Eq.(\ref{SoftMassFormula}).
We have the same number of observable scalar masses as that of species
of scalar fields, e.g., 17 observables in the MSSM.
There are several model-dependent parameters in the RHS of 
Eq.(\ref{SoftMassFormula}) such as
$m_{3/2}^2+V_0$, $\cos\theta$ and so on.
If the number of independent equations is more than that of 
unknown parameters, non-trivial relations
exist among scalar masses.

In our model, the breaking scale $M_I$ is 
estimated as $O(10^{-1} M)$ and so renormalization effects from $M$ to $M_I$ 
are neglected.
We assume that Yukawa couplings among heavy and light fields are 
small enough and the $R$-parity is conserved.
In such a case, we can neglect the effect of extra $F$-term contributions.
In this model, the light fields $\hat{\phi}^k$ equal to just
string states and so there are no mixing terms among heavy and 
light fields in the K\"ahler potential.
As discussed in subsection 3.3, there appear no heavy-light mixing terms
of $O(m_{3/2} M_I)$ if Yukawa couplings among heavy, light and moduli
fields are suppressed sufficiently, i.e., 
$\langle \hat{W}_{Hki} \rangle = O(m_{3/2}/M)$.
At that time, the quantities $\hat{N}_k^l$ and $(\hat{Q}^\alpha)_k^l$ are 
simplified as 
\begin{eqnarray}
\hat{N}_k^l &=& n_k \delta_k^l ,~~~
(\hat{Q}^\alpha)_k^l = q_k^{\alpha} \delta_k^l .
\end{eqnarray}

Under the above assumptions and excellent features,
our soft scalar mass formula is written in a simple form such as,
\begin{eqnarray}
(m^2)_k|_{M_I} &=& m_{3/2}^2 + m_{3/2}^2  n_k \cos^2\theta 
+ \sum_{\hat{\alpha}} g_{\hat{\alpha}}^2
\langle D^{\hat{\alpha}} \rangle q_k^{\hat{\alpha}}
\nonumber\\
&=& m_{3/2}^2\{ 1 + n_k \cos^2\theta
 + {q_k^A \over 12}(5 - 7 \cos^2\theta) \}
\label{m2_k}
\end{eqnarray}
where 
we take $V_0=0$, i.e., $C=1$.
Here we use the formula of $D$-term condensation (\ref{DA-cond}) and the 
following values,
\begin{eqnarray*}
f(q^A)=-12v_1,~~~ f(q^{A} n_\kappa)=12v_1,~~~
f((q^{A})^2)=144v_1. 
\end{eqnarray*}
In this model, the gauge boson mass matrix is diagonalized
for the components of $U(1)_A$ and $U(1)'_3$ up to $m_{3/2}^2/M_I^2$.

In Table 3, we give a ratio $m_k^2/m_{3/2}^2$ at $M_I$ for all 
light species except $G'_{SM}$ singlets
in two extreme cases, $\cos^2 \theta = 0$ and $\cos^2 \theta = 1$.
For $\cos^2 \theta=1$, $L_i$ $(i=3,4,5)$ and $R_j$ $(j=1,4,5)$
fields acquire negative squared masses and 
they could trigger a \lq \lq larger'' symmetry breaking 
including the dangerous charge symmetry breaking.
In addition, we have a strong non-universality of soft masses.  
However, in this model, soft masses are degenerate for squarks 
and sleptons with same quantum numbers under $G_{SM}$
because they have same quantum
numbers under the gauge group $G$ and same modular weights.

We have the following relations by eliminating model-dependent 
parameters,
\begin{eqnarray}
&~& m_{\tilde{Q}_L}^2 = m_{\tilde{Q}_R}^2 = m_H^2 ,
\nonumber\\
&~& m_{\tilde{L}}^2 = m_{\tilde{R}}^2 , 
\nonumber\\
&~& 13m_{\tilde{Q}_L}^2 
= 3m_{\tilde{L}}^2 +5m_{3/2}^2 
\label{Massrelation}
\end{eqnarray}
where the tilde represents scalar components.

On the top of that, the guagino mass $M_{1/2}$ is obtained as \cite{BIM}
\begin{eqnarray}
M^2_{1/2}=3m_{3/2}^2\sin^2 \theta.
\end{eqnarray}
We can use this gaugino mass to obtain a relation not including 
$m_{3/2}$ as 
\begin{eqnarray}
3m_{\tilde{Q}_L}^2 = M^2_{1/2}.
\end{eqnarray}

In the case that the SUSY breaking is induced by the dilaton $F$-term,
there are no modular weight dependence.
Hence we have a more specific relation 
such that
\begin{eqnarray}
&~& 8m_{\tilde{Q}_L}^2 = 3m_{\tilde{L}}^2 .
\label{NewMassrelation}
\end{eqnarray}

Further various contributions should be added at lower energy scales.
For example, the following $D$-term contribution appears after
$U(1)'_i$'s and $SU(2)'$ breakings at $O(m_{3/2})$,
\begin{eqnarray}
(m_D^2)_k|_{m_{3/2}} &=&  -m_{3/2}^2 {q_k^{3'} \over 6\sqrt{2}} 
\rho \cos^2\theta
\label{mD2_k}
\end{eqnarray}
where $\rho=v_2/(v_2+v_3)$.
The ratio $\rho$ takes a value as $\rho \leq 1$ because $v_i \geq 0$. 
To derive Eq.~(\ref{mD2_k}),
we use the formula of $D$-term condensation (\ref{DB-cond}) and the 
following values,
\begin{eqnarray*}
f(q'^{3} n_\kappa)=4\sqrt{2}v_2,~~~
f((q'^{3})^2)=48(v_2+v_3). 
\end{eqnarray*}
Note that the $D$-terms of $Q'^1_D$, $Q'^2_D$ and $T^{3'}_D$ 
do not contribute soft 
scalar masses up to renormalization effects
because $f(q'^{1}_D n_\kappa)=f(q'^{2}_D n_\kappa)=f(q^{3'}_D n_\kappa)=0$.
Here $Q'^1_D$ $(q'^{1}_D)$, $Q'^2_D$ $(q'^{2}_D)$ and
$T^{3'}_D$ $(q^{3'}_D)$ are the diagonal charge operators (charges) 
where the gauge boson mass matrix is diagonalized and they are
constructed as linear combinations of $Q'^1$, $Q'^2$ and $T^{3'}_D$.

In general, original string states are different from the MSSM fields
in string models including $G_{SM}$ \cite{LRmodel}.
The coefficients $R_\kappa^\lambda$ of linear combinations depend on
the VEVs of moduli fields.
A study of flat directions and soft masses
in such a situation has been progressed 
by using explicit models \cite{KKK}.

\section{Remarks on extension of K\"ahler potential}

Here we discuss extensions of our soft mass formula for 
different types of K\"ahler potentials.
At the one-loop level, the dilaton field $S$ and the moduli 
field $T$ are mixed in the K\"ahler potential as 
\begin{eqnarray}
-ln (S+S^*+\Delta(T+T^*))-3 ln (T+T^*).
\end{eqnarray}
In this case we can obtain the same parametrization of soft 
scalar masses as the case without the mixing, i.e. $\Delta(T+T^*)=0$, 
except replacing $\cos^2 \theta$ as 
\begin{eqnarray}
\cos^2 \theta \rightarrow \left[ 1-{(T+T^*)^2\Delta''(T+T^*) \over 
3(S+S^*+\Delta(T+T^*))} \right]\cos^2 \theta
\end{eqnarray}
where $\Delta''(T+T^*)$ is the second derivative of $\Delta(T+T^*)$ 
by $T$.

In general, string models have several moduli fields other than 
one overall moduli field $T$.
In this case, their $F$-terms could contribute on the SUSY breaking 
and one needs more goldstino angles to parametrize these $F$-terms.
For example, we discuss the models with three diagonal 
moduli fields $T_i$ ($i=1,2,3$).
These moduli fields have the following K\"ahler potential:
\begin{eqnarray}
- \sum_i ln (T_i+T_i^*)
\end{eqnarray}
instead of $-3 ln (T+T^*)$ in the case of the overall moduli field.
Here we parametrize their contributions on the SUSY breaking 
as \cite{multiT} 
\begin{eqnarray}
\langle e^{G/2} (K^{T_i}_{T_i})^{-1/2}G^{T_i} \rangle 
=\sqrt 3 Cm_{3/2}e^{i\alpha_{T_i}}\cos \theta \Theta_i
\end{eqnarray}
where $\sum_i \Theta^2_i = 1$.
Using these parameters, $F$-term contributions on soft scalar masses 
are written as 
\begin{eqnarray}
m_{3/2}^2+V_0+3m_{3/2}^2C^2
\sum_i n_{i\kappa}\cos^2 \theta \Theta_i^2
\end{eqnarray}
where $n_{i\kappa}$ is a modular weight of $\phi^\kappa$ for 
the $i$-th moduli field $T_i$.
Similarly $D$-term contributions can be written by the use of 
these parameters.
For example, the $D$-term condensations (\ref{DB-cond}) 
are extended as 
\begin{eqnarray}
g^2_{\hat{\alpha}'} \langle D^{\hat{\alpha}'} \rangle 
= -3 m_{3/2}^2 C^2 \cos^2\theta 
\sum_i \Theta^2_i {\langle \sum_\kappa {q_\kappa^{\hat{\alpha}'}}
 n_{i \kappa} (T+T^*)^{n_\kappa}
 |\phi^\kappa|^2 \rangle  \over
\langle \sum_\kappa (q_\kappa^{\hat{\alpha}'})^2 
(T+T^*)^{n_\kappa} |\phi^\kappa|^2
\rangle}
\end{eqnarray}
where $(T+T^*)^{n_\kappa}$ means 
$\prod_{i=1}^3(T_i+T_i^*)^{n_{i \kappa}}$.

Some orbifold models have complex structure moduli fields $U_i$.
In such models, a K\"ahler potential includes holomorphic parts 
as \cite{holom}
\begin{eqnarray}
{1 \over (T_i+T^*_i) (U_i+U^*_i)}\phi \phi'.
\end{eqnarray}
We can extend our formula into these models.
These holomorphic parts are important for mixing of fields.
Further they could originate the $\mu$-term with 
a suitable order naturally.

The K\"ahler potential can receive radiative corrections and
be modified by non-perturbative effects.
Our approach is generic and basically available to other types
of K\"ahler potential although one might need more complicated 
parametrization than (\ref{G^S}) and (\ref{G^T}).

\section{Conclusions and Discussions}

We have derived the formula of soft SUSY breaking scalar masses from 
the effective SUGRA derived from 4-D string models 
within a more generic framework.
The gauge group contains extra gauge symmetries including
the anomalous $U(1)$ some of which are broken at a higher energy scale.
The breakings are related to the flat direction breakings 
in the SUSY limit.
It is supposed that there are two types of matter multiplets classified by 
supersymmetric fermion mass, i.e., heavy fields and light ones.
The physical scalar fields are, in general, linear combinations of
original fields corresponding to massless states in string models. 

The mass formula contains the effects of extra gauge symmetry breakings, 
i.e., $D$-term and extra $F$-term contributions,
particle mixing effects and heavy-light mass mixing effects.
The $D$-term contributions to soft scalar masses are 
parameterized in terms of three types of new parameter in addition to
the goldstino angle, gravitino mass and vacuum energy.
These contributions, in general, are sizable.
In particular, $D$-term contribution of $U(1)_A$ survives even 
in the case of the dilaton dominant SUSY breaking.
The $D$-term contributions for anomaly-free $U(1)$ symmetries vanish 
if the fields developing VEVs have the same modular weight.
Extra $F$-term contributions are neglected in the case that Yukawa couplings
among Nambu-Goldstone, heavy and light fields are suppressed and
the $R$-parity is conserved. 
In the case that there exist heavy-light mixing terms in the K\"ahler
potential, the extra contributions can appear after
the diagonalization of scalar mass terms in the presence of
heavy-light mass mixings of $O(m_{3/2} M_I)$.

We have discussed the degeneracy and the positivity of squared scalar masses
in special cases that there are neither particle mixing in the K\"ahler
potential, heavy-light mass mixing effects 
nor extra $F$-term contributions.
We find that the $F$-term contribution from the difference among
modular weights and the $D$-term contribution to scalar masses
can destroy universality among scalar masses at $M$.
This non-degeneracy endangers the discussion of the suppression
of FCNC process.
On the other hand, the difference among $U(1)$ charges
is crucial for the generation of fermion mass hierarchy.
It seems to be difficult to make two discussions compatible.
As a byway, we can take a model that the fermion mass hierarchy is 
generated due to non-anomalous $U(1)$ symmetries
and SUSY is broken by the dilaton $F$-term condensation.
For example, it is supposed that anomalies from contributions of
the MSSM matter fields are canceled out 
by those of extra matter fields in such a model.
Further ``stringy'' symmetries are also useful for fermion mass generation 
leading to degenerate soft scalar masses \cite{texture2}, 
because these symmetries do not induce $D$-terms.

Many fields could acquire negative squared masses and 
they could trigger a \lq \lq larger'' symmetry breaking 
including the dangerous color and/or charge symmetry breaking.
This type of symmetry breaking might be favorable in the case where 
$G'_{SM}$ is a large group like a grand unified group.
These results might be useful for model building.

We have derived specific scalar mass relations by taking an explicit
string model.
It is expected that such relations can be novel probes to select
a realistic string model since they are model-dependent.

The moduli fields have a problem in string cosmology
because their masses are estimated as of $O(m_{3/2})$
and they weakly couple with 
the observable matter fields, i.e. through the gravitational 
couplings \cite{cosmo}.
They decay slowly to the observable matter fields.
That makes the standard nucleosynthesis dangerous.
In our model, some linear combinations of $S$, $T$ and 
other fields like $X$ remain light
whose $F$-terms are of $O(m_{3/2} M)$ and break the SUSY.
It is supposed that the couplings between such fields and observable
fields are strongly suppressed to guarantee the stability 
of the weak scale.
Such a problem have to be considered for the light linear combinations, too.

\section*{Acknowledgments}
The authors are grateful to I.~Joichi, T.~Komatsu, J.~Louis, H.~Nakano, 
D.~Suematsu and M.~Yamaguchi 
for useful discussions.

\appendix

\section{K\"ahler Potential and its Derivatives in String Models}
\label{app:A}

The K\"ahler potential $K$ in $Z_N$ orbifold models is given as 
\cite{ST-SG,OrbSG,OrbSG2}
\begin{eqnarray}
   K &=& -ln(S + S^*) - 3ln(T + T^*)
\nonumber\\ 
&~&  + \sum_{\kappa} (T + T^*)^{n_\kappa}|\phi^\kappa|^2 + \cdots
\label{ap-K}
\end{eqnarray}
in the case of overall moduli.
The derivatives of $K$ are given as
\begin{eqnarray}
   K_S^S &=& {1 \over (S+S^*)^2},~~~K_S^T =0,~~~K_S^\kappa =0,
\nonumber\\
   K_T^T &=& {3 \over (T+T^*)^2} + \sum_\kappa n_\kappa(n_\kappa -1)
             (T + T^*)^{n_\kappa-2}|\phi^\kappa|^2 ,
\nonumber\\ 
   K_T^\lambda &=& n_\lambda (T + T^*)^{n_\lambda-1}\phi^\lambda,~~~
   K_\kappa^\lambda = (T + T^*)^{n_\kappa-1}\delta^\lambda_\kappa .
\nonumber 
\end{eqnarray}
The determinant $K_I^J$ is calculated as
\begin{eqnarray}
   \Delta &\equiv& {\rm det}K_I^J
\nonumber\\
 &=& {3 \prod_\lambda (T+T^*)^{n_\lambda} \over (S+S^*)^2(T+T^*)^2}
               \{ 1 - \sum_\kappa {n_\kappa \over 3}
             (T + T^*)^{n_\kappa}|\phi^\kappa|^2 \} .
\label{det}
\end{eqnarray}
The inverses $(K_I^J)^{-1}$ are given as
\begin{eqnarray}
   (K_S^S)^{-1} &=& (S+S^*)^2,~~~(K_S^T)^{-1} =0,~~~
    (K_S^\kappa)^{-1} =0,
\nonumber\\
   (K_T^T)^{-1} &=& 
   {\prod_{\kappa}(T+T^*)^{n_\kappa} \over (S+S^*)^2\Delta}
\nonumber\\ 
   &=& {(T+T^*)^2 \over 3}\{1 + \sum_\kappa {n_\kappa \over 3}
             (T + T^*)^{n_\kappa}|\phi^\kappa|^2\} + O(|\phi|^4) ,
\nonumber\\ 
   (K_\kappa^T)^{-1} &=& 
- {n_\kappa (T + T^*) \phi^*_\kappa \over
   3 - \sum_\lambda n_\lambda (T + T^*)^{n_\lambda}|\phi^\lambda|^2}
\nonumber\\ 
 &=& - {n_\kappa \over 3}(T + T^*) \phi^*_\kappa \{1 + 
\sum_\lambda {n_\lambda \over 3}(T + T^*)^{n_\lambda}|\phi^\lambda|^2 \}
+ O(\phi^*|\phi|^4) ,
\nonumber\\ 
   (K_\kappa^\lambda)^{-1} &=& 
{3(T + T^*)^{-n_\kappa} \delta_\kappa^\lambda \over
   3 - \sum_\lambda n_\lambda (T + T^*)^{n_\lambda}|\phi^\lambda|^2}
\nonumber\\ 
&=& (T + T^*)^{-n_\kappa} \delta_\kappa^\lambda 
+ O(|\phi|^2) 
\nonumber
\end{eqnarray}
where $\phi$ represents scalar field of matter multiplet.

\newpage

\section*{Table Captions}

\renewcommand{\labelenumi}{Table~\arabic{enumi}}
\begin{enumerate}
\item $U(1)$ charge generators in terms of $E_8 \times E_8'$ lattice
vectors.
An anomalous $U(1)$ charge $Q^A$ is defined as $Q^A \equiv Q_5-Q_6$.
We denote the third component of generators of $SU(2)'$ as $T^{3'}$
and the number in the seventh column represents the eigenvalue
of $2T^{3'}$ for the field component with VEV.

\item The modular weights and broken $U(1)$ charges for the light 
scalar fields and the scalar fields with large VEVs.

\item The particle contents and the ratios of $m_k^2/m_{3/2}^2$.
We refer to the chiral multiplets as $Q_L$ for left-handed quarks,
$Q_R$ right-handed quarks, $H$ Higgs doublets,
$L$ for left-handed leptons and $R$ for right-handed leptons.
The fields $L'$, $\bar{L}'$ and $R'$, $\bar{R}'$ are
extra $SU(2)_L$ and $SU(2)_R$ doublets, respectively.
\end{enumerate}

\newpage

\begin{center}
{\Large Table 1} 
\end{center}
\begin{eqnarray*}
Q_1 &=& 6(1,1,1,0,0,0,0,0)(0,0,0,0,0,0,0,0)'\\
Q_2 &=& 6(0,0,0,1,-1,0,0,0)(0,0,0,0,0,0,0,0)'\\
Q_3 &=& 6(0,0,0,0,0,1,1,0)(0,0,0,0,0,0,0,0)'\\
Q_4 &=& 6(0,0,0,0,0,0,0,1)(0,0,0,0,0,0,0,0)'\\
Q_5 &=& 6(0,0,0,0,0,0,0,0)(1,0,0,0,0,0,0,0)'\\
Q_6 &=& 6(0,0,0,0,0,0,0,0)(0,1,0,0,0,0,0,0)'\\
Q_7 &=& 6(0,0,0,0,0,0,0,0)(0,0,1,1,0,0,0,0)'
\end{eqnarray*}

~~\

\begin{center}
{\Large Table 2} 
\end{center}

\begin{center}
\begin{tabular}{|c|c|c|c|c|c|c|}\hline
 String states & $n_k$ & $Q'^1$ & $\sqrt{3}Q'^2$ 
& $\sqrt{2}Q'^3$ & $Q^A$ & $2T^{3'}$\\
\hline
$Q_L$ & $-1$ & 6 & 0 & $-6$ & 0 & 0 \\ 
$Q_R$ & $-1$ & $-6$ & 0 & $-6$ & 0 & 0 \\ 
$H$ & $-1$ & 0 & 0 & 12 & 0 & 0 \\ 
$L$ $(L_4)$ & $-2$ & $-2$ & 0 & $-2$ & 4 & 0 \\ 
$R$ $(R_5)$ & $-2$ & 2 & 0 & $-2$ & 4 & 0 \\ 
$L'$ $(L_3)$ & $-2$ & $-2$ & 0 & $-2$ & 4  & 0 \\ 
$\bar{L}'$ $(L_5)$ & $-2$ & $-2$ & 0 & $-2$ & 4 & 0 \\ 
$R'$ $(R_4)$ & $-2$ & 2 & 0 & $-2$ & 4 & 0 \\ 
$\bar{R}'$ $(R_1)$ & $-2$ & 2 & 0 & $-2$ & 4 & 0 \\ \hline
$u$ & $-1$ & 0 & 0 & 0 & $-12$ & 0 \\ 
$Y$ & $-3$ & 0 & 0 & $-8$ & 4 & 0 \\ 
$S_1$ & $-2$ & 0 & 0 & $-8$ & $-8$ & 0 \\ 
$D'_3$ & $-2$ & 6 & $-6$ & 4 & $-2$ & 1 \\ 
$D'_4$ & $-2$ & 6 & 6 & 4 & $-2$ & $-1$ \\ 
$D'_5$ & $-2$ & $-6$ & $-6$ & 4 & $-2$ & 1 \\ 
$D'_6$ & $-2$ & $-6$ & 6 & 4 & $-2$ & $-1$ \\ \hline
\end{tabular}
\end{center}

\newpage

\begin{center}
{\Large Table 3} 
\end{center}

\begin{center}
\begin{tabular}{|c|c|c|c|c|}\hline
 & Rep. & $q_k^A$ & \multicolumn{2}{|c|}{$m_k^2/m_{3/2}^2|_{M_I}$}\\
 & & & $\cos^2 \theta=0$ & $\cos^2 \theta=1$\\ \hline 
U-sec. & $Q_L$ $(3,2,1)$  & 0 & 1 & 0 \\
       & $Q_R$ $(\bar 3,1,2)$ &  0 & 1 & 0 \\
       & $H$ $(1,2,2)$  & 0 & 1 & 0 \\ \hline
T-sec. & $L$ $(1,2,1)$  & 4  &  8/3  & $-5/3$ \\ 
$(N_{OSC}=0)$  & $R$ $(1,1,2)$  & 4  &  8/3  & $-5/3$ \\ 
      & $L'$ $(1,2,1)$  & 4  &  8/3  & $-5/3$ \\ 
      & $\bar{L}'$ $(1,2,1)$  & 4  &  8/3  & $-5/3$ \\ 
      & $R'$ $(1,1,2)$  & 4  &  8/3  & $-5/3$ \\ 
      & $\bar{R}'$ $(1,1,2)$  & 4  &  8/3  & $-5/3$ \\ \hline
\end{tabular}
\end{center}

\end{document}